\documentclass[aps,prb,reprint,superscriptaddress,showpacs,showkeys]{revtex4-2}%
\usepackage{amsfonts}
\usepackage{amssymb}
\usepackage{amsmath}
\usepackage{graphicx}
\usepackage{dcolumn}
\usepackage{bm}
\usepackage{units}
\usepackage{multirow}
\usepackage{CJKutf8}
\usepackage{subfigure}
\usepackage{color}%
\usepackage{url}
\usepackage[colorlinks,linkcolor=red,anchorcolor=green,citecolor=blue]{hyperref}
\providecommand{\U}[1]{\protect\rule{.1in}{.1in}}
\usepackage{array}
\newcommand{\PreserveBackslash}[1]{\let\temp=\\#1\let\\=\temp}
\newcolumntype{C}[1]{>{\PreserveBackslash\centering}p{#1}}
\newcolumntype{R}[1]{>{\PreserveBackslash\raggedleft}p{#1}}
\newcolumntype{L}[1]{>{\PreserveBackslash\raggedright}p{#1}}
\begin{document}
\title{Unified First-Principles Study of the Anomalous Hall Effect Based on Exact Muffin-Tin Orbitals}
\author{Lei Wang (\begin{CJK}{UTF8}{gbsn}王蕾\end{CJK})}
\affiliation{Center for Spintronics and Quantum Systems, State Key Laboratory for Mechanical Behavior of Materials, Xi'an Jiaotong University, No.28 Xianning West Road Xi'an, Shaanxi, 710049, China}
\author{Tai Min}
\email{tai.min@mail.xjtu.edu.cn}
\affiliation{Center for Spintronics and Quantum Systems, State Key Laboratory for Mechanical Behavior of Materials, Xi'an Jiaotong University, No.28 Xianning West Road Xi'an, Shaanxi, 710049, China}
\author{Ke Xia}
\email{kexia@bnu.edu.cn}
\affiliation{Beijing Computational Science Research Center, Beijing, 100193, China}

\date{\today}

\begin{abstract}
Based on the exact muffin-tin orbitals (EMTOs), we developed a first-principles method to calculate the current operators and investigated the anomalous Hall effect in bcc Fe as an example, with which we successfully separated the skew scattering contribution from the side jump and intrinsic contributions by fitting the scaling law with the introduction of sparse impurities. By investigating the temperature dependence of the anomalous Hall effect in bulk Fe, we predicted a fluctuated anomalous Hall angle as a function of temperature when considering only phonons, which, in the future, can be measured in experiments by suppressing magnon excitation, e.g., by applying a high external magnetic field.
\end{abstract}

\maketitle

\section{Introduction} 
The anomalous Hall effect discovered by Hall \cite{Hall1880} is one of the most important effects in the field of spintronics and refers to the generation of a charge current normal to the primary electrical current and the magnetization in a ferromagnetic conductor \cite{PhysRev.95.1154,PhysRevLett.88.207208,RevModPhys.82.1539,Kondo01041962,natm3.4,PhysRevLett.92.166601,Fang92,PhysRevLett.92.037204} in the absence of an external magnetic field. The origins of the anomalous Hall effect are mainly ascribed to the intrinsic Berry curvature \cite{PhysRevB.53.7010,PhysRevB.59.14915} and extrinsic skew scattering \cite{SMIT1955877,SMIT195839} and side jump \cite{PhysRevB.2.4559} mechanisms.

However, compared to the well-studied intrinsic contribution obtained by Berry curvature calculations \cite{PhysRevLett.92.037204,PhysRevLett.93.206602,PhysRevB.76.195109,PhysRevB.99.224416}, the extrinsic contributions are much more complicated due to the multiple impurities in experiments, such as defects \cite{jap1.353821}, phonons \cite{PhysRevLett.115.076602,2012JPSJ,PhysRevB.76.195309}, alloys \cite{PhysRevLett.107.106601,PhysRevLett.105.266604}, amorphous disorder \cite{PhysRevB.90.214410} and surface roughness \cite{Grigoryan2016Scaling}. Thus, more work should be done to study the extrinsic contributions to the anomalous Hall effect. For the extrinsic contributions, the side jump contribution is independent of the scattering strength and disorder density \cite{PhysRevB.83.125122} and is treated as constant and as being entangled with the intrinsic contribution in experiments \cite{PhysRevLett.114.217203,PhysRevB.85.220403}. Therefore, skew scattering, an adjustable contribution, has more possibilities for applications and needs to be studied individually by separating it from the intrinsic and side jump contributions.

Recently, a first-principles calculation method using the Landauer-Büttiker formalism based on linear muffin-tin orbitals (LMTOs) was reported to be able to calculate the full current operators for disordered systems \cite{PhysRevB.73.064420,PhysRevB.77.184430,Wang.2016,PhysRevB.97.214415,PhysRevB.99.144409,PhysRevB.99.134427} and was successfully used to investigate spin current-related physical issues, such as the spin Hall effect and spin diffusion length. Furthermore, improved EMTOs have been reported over the years \cite{andersen1995exact,PhysRevB.64.014107,VITOS200024,PhysRevB.71.094415} and been used to study alloys \cite{PhysRevLett.87.156401,PhysRevB.87.075144}, surfaces/interfaces \cite{VITOS1998186,PhysRevB.102.035405} and magnetic tunnel junctions \cite{PhysRevB.100.075134}. Based on the above progress, in this work, we reproduce the full current operators \cite{PhysRevB.77.184430,Wang.2016,PhysRevB.97.214415,PhysRevB.99.144409} using the EMTOs and apply them to study the anomalous Hall effect in Fe. To distinguish the extrinsic contributions to the anomalous Hall effect, we first calculate the anomalous Hall conductivity for bcc Fe by introducing sparse impurities, such as C, Cr, Cu, Pd, Ag and Pt, into Fe at zero temperature and investigate the corresponding scaling law to separate the skew scattering contribution from the side jump and intrinsic contributions. We also study the temperature-dependent anomalous Hall angle with magnons and phonons and find an unexpected fluctuation with only phonons. In this sense, this fluctuated anomalous Hall angle can be measured in future experiments by suppressing magnon excitation using various methods, such as applying a high external magnetic field.

\begin{figure}[tp]
	\includegraphics[width=\columnwidth]{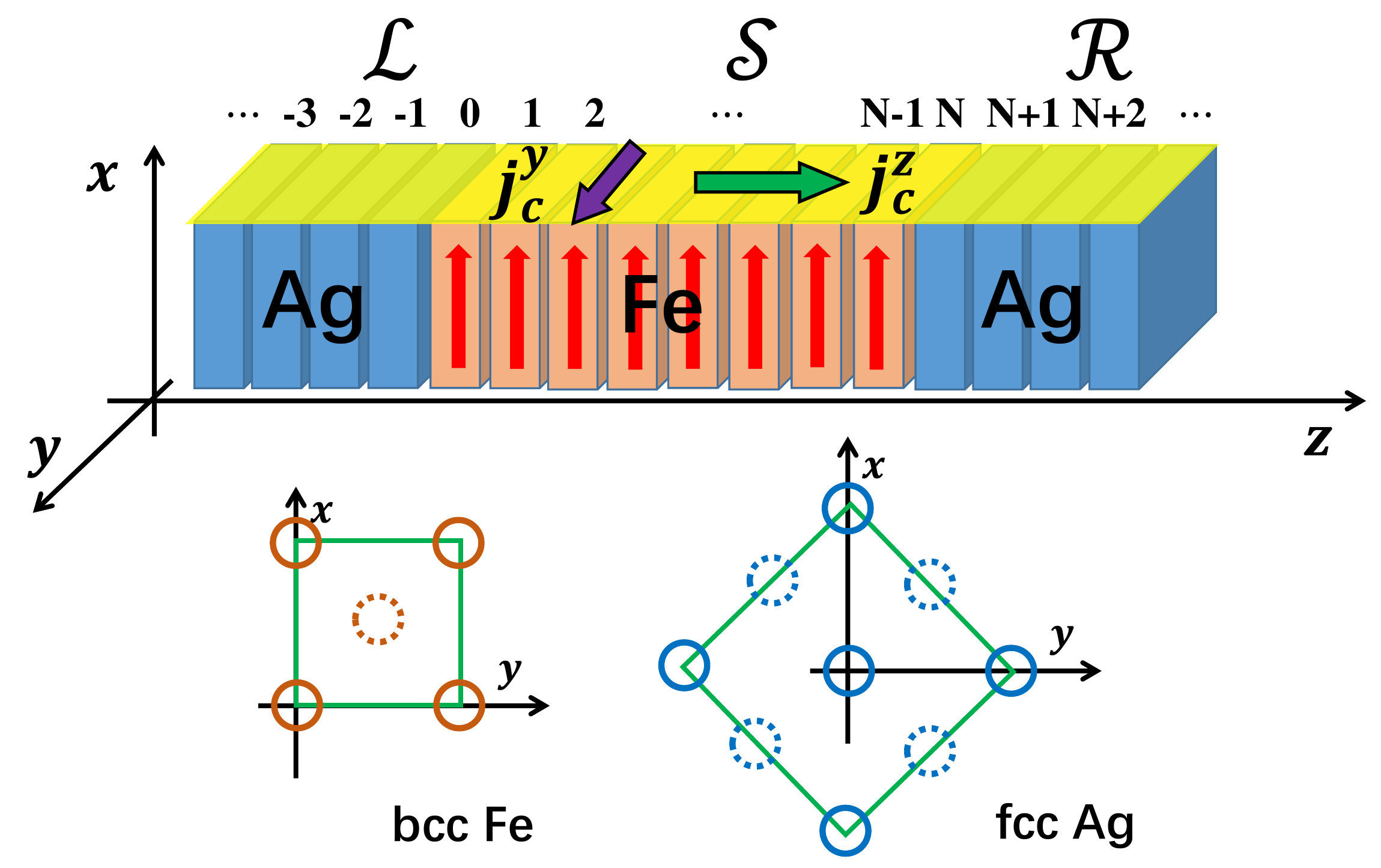}
	\caption{Scattering model with fcc Ag leads and bcc Fe in between. The transport direction ($z$) is along the [001] direction of the Fe lattice. The Ag lattice is rotated by $45^{\circ}$ to match the Fe lattice. With magnetization along the $x$-axis, the generated anomalous Hall current flows along the $y$-axis.}
	\label{trans}
\end{figure}

\section{Model and method}

Within the frame of the Landauer-Büttiker transport theory \cite{PhysRevB.31.6207,PhysRevB.45.1347,PhysRevB.44.10637,datta_1995,Imry_2002}, the structure of the calculated system in this paper is constructed by two leads and one scattering region as shown in Fig.~\ref{trans}, where the two leads ($\mathcal{L}$ and $\mathcal{R}$) are semi-infinite crystallines, used to inject electrons from the left or right lead to the scattering region ($\mathcal{S}$), respectively. And following Ref.~\cite{PhysRevB.73.064420,PhysRevB.97.214415}, the scattering wave functions of the whole scattering region ($\mathcal{S}$) are calculated layer by layer using the wave-function matching method, which can be used to calculate the currents \cite{PhysRevB.73.064420,PhysRevB.97.214415,PhysRevB.77.184430,Wang.2016,PhysRevB.99.144409}, and thereby the anomalous Hall effect as follows.

\subsection{Scattering wave functions}
\label{swf} 

The screened Korringa-Kohn-Rostoker (KKR) equation for the EMTOs basis can be written as \cite{andersen1995exact,vitos2007computational,PhysRevB.100.075134}
\begin{eqnarray}
\sum_R a_{R'}\left[\hat{S}_{R'R}-\delta_{R'R}\hat{D}_R(\epsilon)\right]\mathbf{V}_R=0,
\end{eqnarray}
where $a_{R'}$ is the hard sphere radius of atom $R'$, $\mathbf{V}_R$ is the corresponding expansion coefficient vector for the EMTOs basis, and $\hat{S}_{R'R}$ and $\hat{D}_R(\epsilon)$ are the slope matrices and logarithmic derivative matrices at a defined energy $\epsilon$, respectively, which can be obtained by the EMTO-CPA self-consistent code \cite{PhysRevB.100.075134,emto,PhysRevLett.87.156401,PhysRevB.71.094415}.

For a transport system with lateral ($x$-$y$ plane) periodic boundary conditions, as shown in Fig.~\ref{trans}, the KKR equation above can be transformed into a layer-resolved representation as follows:
\begin{eqnarray}
-\hat{S}_{I,I-1}^{\mathbf{k}_{\parallel}}\mathbf{V}_{I-1}+(\hat{D}_{I}(\epsilon)-\hat{S}_{I,I}^{\mathbf{k}_{\parallel}})\mathbf{V}_{I}-\hat{S}_{I,I+1}^{\mathbf{k}_{\parallel}}\mathbf{V}_{I+1}=0,\ \ \ \ 
\label{kkr}
\end{eqnarray}
with
\begin{eqnarray}
\hat{S}_{I,J}^{\mathbf{k}_{\parallel}}=\sum_{\mathbf{T}\in\{\mathbf{T}_{IJ}\}}\hat{S}(\mathbf{T}) e^{i\mathbf{k}_{\parallel}\cdot\mathbf{T}},
\end{eqnarray}
where $I$ and $J$ denote the layer index, $\{\mathbf{T}_{IJ}\}$ represents the vectors that connect one lattice site in the $I$th layer with lattice sites in the $J$th layer, and $\mathbf{k}_{\parallel}$ denotes the reciprocal lattice inside the lateral Brillouin zone.

The form of Eq.~(\ref{kkr}) is the same as that of the equation of motion of electrons in the  wave-function matching method \cite{PhysRevB.44.8017,PhysRevB.73.064420,PhysRevB.97.214415}. Thus, we can follow Ref.~\cite{PhysRevB.73.064420,PhysRevB.97.214415} to obtain the scattering wave functions of the whole system as
\begin{eqnarray}
	\Psi\equiv\left(\begin{array}{c}
		\mathbf{V}_0 \\
		\mathbf{V}_1 \\
		\mathbf{V}_2 \\
		\vdots \\
		\mathbf{V}_N \\
		\mathbf{V}_{N+1}
	\end{array}\right)=\left(\tilde{\mathbf{D}}-\tilde{\mathbf{S}}\right)^{-1}\times
\left(\begin{array}{c}
	\mathbf{\tilde{V}}_0 \\
	0 \\
	0 \\
	\vdots \\
	0 \\
	0
\end{array}\right),
\label{ds}
\end{eqnarray}
where $\tilde{\mathbf{D}}$ and $\tilde{\mathbf{S}}$ are block tridiagonal matrices that contain all $\hat{D}_{I}(\epsilon)$ and $\hat{S}_{I,J}^{\mathbf{k}_{\parallel}}$ of the whole system, respectively. One may notice that, except directly putting the  $\hat{D}_{I}(\epsilon)$ and $\hat{S}_{I,J}^{\mathbf{k}_{\parallel}}$ in the corresponding location in  $\tilde{\mathbf{D}}$ and $\tilde{\mathbf{S}}$ matrix, the slope matrices in the leads ($\tilde{S}_{0,0}^{\mathbf{k}_{\parallel}}$, $\tilde{S}_{N+1,N+1}^{\mathbf{k}_{\parallel}}$) and the injecting wave function from $\mathcal{L}$ lead ($\mathbf{\tilde{V}}_0$) are re-normalized by the boundary condition in the leads~\cite{PhysRevB.73.064420,PhysRevB.97.214415}, accordingly.

\subsection{Current operators} 
\label{oper}
With the scattering wave function $\Psi$ from Sec.~\ref{swf} and projecting it into the basis of every atom $R$ inside the scattering region ($\mathcal{S}$), the atomic scattering wave functions for all atoms can be obtained and marked as $\Psi_{R}$. And therefore, the corresponding local charge density will be $n_R=\langle\Psi_R\vert\Psi_R\rangle$. It is known that, the time derivative of the charge density on a single atom $R$ comes from the charge current from the surrounded atoms $R'$, reads, 
\begin{eqnarray}
\frac{\partial n_R}{\partial t}=\sum_{R'}J_{RR'},
\end{eqnarray}
where $J_{RR'}$ is the local charge current from the atom $R'$ to $R$. Considering a two-atom system, the Schr\"{o}dinger equation can be written as
\begin{eqnarray}
\partial_t\left(\begin{array}{c}
\Psi_R \\
\Psi_{R'}
\end{array}\right)=\frac{1}{i\hbar}\left(
\begin{array}{cc}
\hat{\mathcal H}_{RR} & \hat{\mathcal H}_{RR'} \\
\hat{\mathcal H}_{R'R} & \hat{\mathcal H}_{R'R'}
\end{array}\right)\left(\begin{array}{c}
\Psi_R \\
\Psi_{R'}
\end{array}\right),
\end{eqnarray}
therefore, we can obtain that,
\begin{eqnarray}
\begin{split}
\frac{\partial n_R}{\partial t}&=\langle\partial_t\Psi_R\vert\Psi_R\rangle+\langle\Psi_R\vert\partial_t\Psi_R\rangle \\
&=\frac{1}{i\hbar}\left[\langle\Psi_R\vert\hat{\mathcal H}_{RR'}\vert\Psi_{R'}\rangle-\langle\Psi_{R'}\vert\hat{\mathcal H}_{R'R}\vert\Psi_{R}\rangle\right].
\end{split}
\end{eqnarray}
Then, the local charge current from atom $R'$ to atom $R$ is given by \cite{PhysRevB.65.125101,PhysRevB.77.184430,Wang.2016,PhysRevB.99.144409,PhysRevB.97.214415,PhysRevB.102.035405}
\begin{eqnarray}
\begin{split}
J_{RR'}=\frac{1}{i\hbar}\left[\langle\Psi_R\vert\hat{\mathcal H}_{RR'}\vert\Psi_{R'}\rangle-\langle\Psi_{R'}\vert\hat{\mathcal H}_{R'R}\vert\Psi_{R}\rangle\right].
\end{split}
\end{eqnarray}
In this sense, the electronic transport properties between any two atoms of the whole system can be estimated and used for the study of the anomalous Hall effect.

\subsection{Anomalous Hall effect}
By calculating all local charge currents $J_{RR'}$ between any two atoms inside the scattering region ($\mathcal{S}$) and projecting the current density in the longitudinal and transverse directions \cite{Wang.2016,PhysRevB.99.144409}, we obtain the primary charge current density $j_c^z$ and the Hall current density $j_c^y$ when $\mathbf{m}\parallel x$, respectively. Thus, the anomalous Hall angle is then given by $\Theta^{\rm AH}=j_c^y/j_c^z$.
\begin{figure}[t]
	\includegraphics[width=\columnwidth]{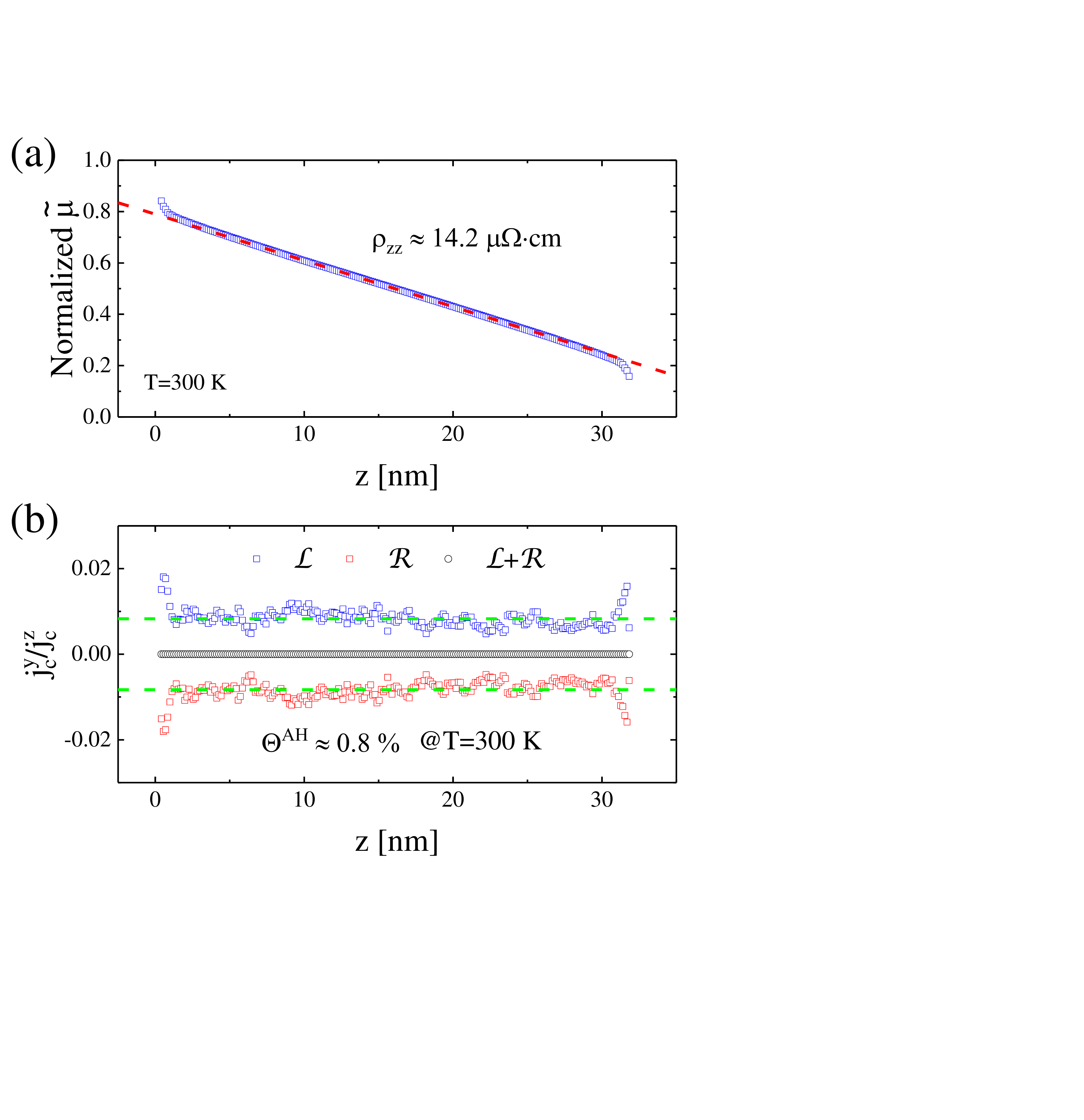}
	\caption{(a) Calculated normalized chemical potential at room temperature (T=300 K) as a function of the atomic position along the transport direction ($z$). (b) Corresponding anomalous Hall current in the entire scattering region, where ``$\mathcal{L}$'' and ``$\mathcal{R}$'' represent the calculated results with injecting electrons from left and right leads, respectively.}
	\label{aha}
\end{figure}

Moreover, the localized scattering wave functions $\Psi_R^{\pm}$ can be solved by injecting electrons from both leads, where $\pm$ denotes the right-going and left-going electrons. Thus, the nonequilibrium density of states can be obtained as $n_R^{\pm}=\langle\Psi_R^{\pm}|\Psi_R^{\pm}\rangle V_b$, induced by the small voltage $V_b$. In this way, we can define a normalized chemical potential $\tilde{\mu}_R=n^+_R/(n^+_R+n^-_R)$ and project them to the longitudinal direction ($z$) to obtain $\tilde{\mu}(z)$. Using the Ohm's law
\begin{eqnarray}
	j_c^z=-\frac{1}{\rho_{zz}}\frac{\partial \tilde{\mu}(z)}{\partial z}V_b,
\end{eqnarray}
the longitudinal resistivity $\rho_{zz}$ can also be calculated. Thus, the corresponding anomalous Hall conductivity will be $\sigma^{\rm AH}=\Theta^{\rm AH}/\rho_{zz}$.

Typically, at room temperature (T=300 K) with both phonons and magnons contributions, we plot the calculated normalized chemical potential ($\tilde{\mu}$) and the normalized anomalous Hall current ($j_c^y/j_c^z$) as a function of the $z$ coordinate of Fe in Fig.~\ref{aha} (a) and (b), respectively. Except for the distortion around the Ag$|$Fe interfaces, both the normalized chemical potential and the anomalous Hall current agree with our previous statements. With a simple linear fitting, we have the resistivity $\rho_{zz}\simeq14.2~\mu\Omega\cdot cm$ and anomalous Hall angle $\Theta^{\rm AH}\simeq0.8\%$ for Fe at room temperature.

Furthermore, we should know that, the anomalous Hall effect was treated as a bulk Fermi sea effect for long time, which needs to calculate all the contributions under Fermi energy \cite{PhysRevLett.92.037204,RevModPhys.82.1539,PhysRevB.99.224416}. And this concept is apparently different with the spirit of the Landauer-Büttiker transport theory \cite{PhysRevB.31.6207,PhysRevB.45.1347,PhysRevB.44.10637}, where only the contribution at Fermi energy is necessary for the transport calculations. However, it had recently been proved that, for the ``nonquantized part'' of the intrinsic Hall conductivity, the integration of the Berry curvature over the entire Fermi see can be equivalently reduced to an alternative integration on the Fermi surface \cite{PhysRevLett.93.206602} and confirmed by individual first principle calculations \cite{PhysRevB.76.195109}. And to verify these concepts using our approach, we calculate the anomalous Hall current with injecting electrons from both left and right leads as shown in Fig.~\ref{aha} (b). It can be seen that, the anomalous Hall currents induced by injecting electrons from left and right leads will annihilate each other (black circles) at finite temperature, indicating that the contribution of the Fermi sea is negligible, which is consistent with the spirit of the conventional Landauer-Büttiker transport theory. Thus, by taking the advantage of the Landauer-Büttiker approach on the disorders calculations, the extrinsic mechanisms can be taken into account easily, which may supply a comprehensive understanding of the origin of the anomalous Hall effect.

\subsection{Computational details} 

The scattering geometry under study is shown in Fig.~\ref{trans}, where the scattering region (Fe) is sandwiched by two semi-infinite crystalline Ag leads. The transport direction ($z$-axis) is set to be along the $[001]$ direction of the bcc Fe lattice. The scattering region is sufficiently long ($\sim 31.6$~nm with 220 atomic layers) such that the influence of the interfaces is negligible around the center. In the $x$-$y$ plane, we use 5$\times$5 lateral bcc supercells with periodic boundary conditions. For the sake of convenience, the fcc lattices in the Ag leads are rotated by $45^{\circ}$ with a 0.65\% stretch of the lattice constant to match the Fe lattice at the interfaces (see the lower panel of Fig.~\ref{trans}). Since the Fe lattice remains in its natural structure and lattice constant $a_{\mathrm{Fe}}$ is $2.87$~\AA, the transport properties extracted from the center of the scattering region are bulk properties and free from the small lattice stretch in the leads.

In general, there are three typical kinds of disorders in our work, impurity, phonon and magnon, and these disorders are introduced independently, thus we can study the disorder effect individually or with arbitrary combination of those three typical kinds of disorders. Among these three types of disorders, the impurity is much easier to generate, in which, only the random replacement by different types of atoms on every atomic position is needed under the control of the established concentration of the materials \cite{PhysRevB.73.064420}. And for the phonon, we employ the static limit by introducing a random displacement to each atom \cite{ziman1960,jap1.3638694,PhysRevB.84.014412,PhysRevB.91.220405}. The displacements are assumed to satisfy a Gaussian distribution with a temperature-dependent variance estimated by the Debye model~\cite{jap1.3638694}, which has been demonstrated to be able to recover the temperature dependences of the resistivity, spin diffusion length and spin Hall effect observed in experiments~\cite{jap1.3638694,PhysRevB.84.014412,PhysRevB.91.220405,Wang.2016,PhysRevB.99.144409}. In the calculation, the Debye temperature of Fe is $T_{D}=470\ K$~\cite{Kittel2005}. For the magnons, following Ref.~\cite{PhysRevB.84.014412,PhysRevB.91.220405}, a random series of spherical coordinates $\vartheta_i$ and $\phi_i$ is generated to map the instantaneous static local magnetization configuration. Note that Gaussian distributions are applied to $\vartheta_i$ under the condition $\langle\cos\vartheta_i\rangle\simeq M(T)/M_s$, which performs well in resistivity calculations \cite{PhysRevB.84.014412,PhysRevB.91.220405}.

\begin{figure}[bp]
	\includegraphics[width=\columnwidth]{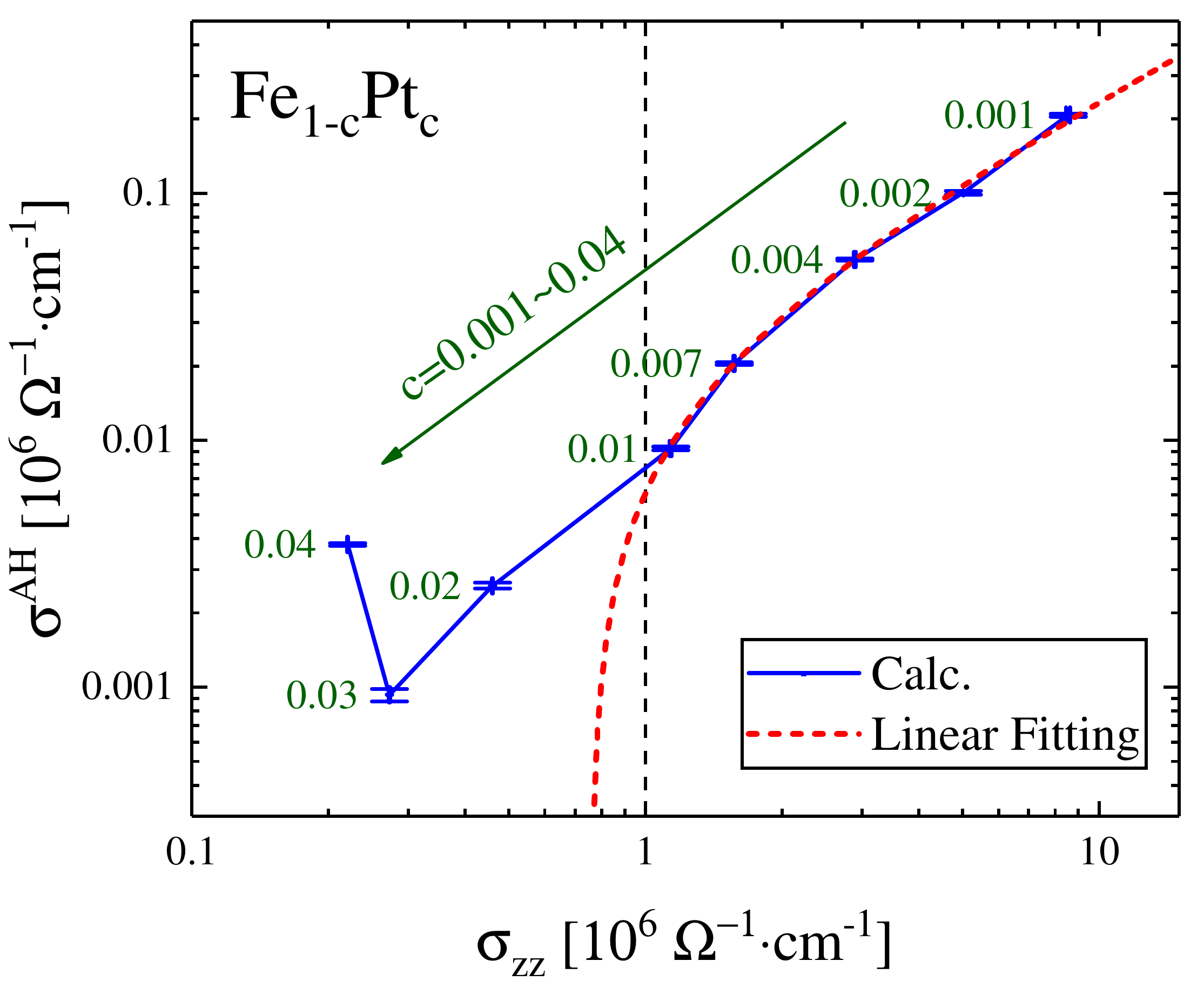}
	\caption{Anomalous Hall conductivity $\sigma^{\rm AH}=\Theta^{\rm AH}/\rho_{zz}$ of Fe$_{1-c}$Pt$_c$ as a function of the longitudinal conductivity $\sigma_{zz}=1/\rho_{zz}$, where $c\in\{0.001\sim0.04\}$ is the corresponding concentration of Pt with labels for all calculated points. The red dash line is for the linear fitting when $c<0.01$ and the vertical dash line separate the nonlinear and linear region.}
	\label{linear}
\end{figure}

And in our approach, the lateral supercells can be treated as a super lattice, however, as the electrons have a limited scattering length under disorder conditions, the super lattice effects will be very small and can be suppressed by the average of large enough configurations. Thus, in this work, we use 10 configurations for the calculations with impurities and 30 configurations for the temperature calculations and discretize the lateral Brillouin zone into an 80$\times$80 mesh to converge the outputs.

\section{Results and analysis}
\subsection{Scaling law} 
The scaling law is generally used to distinguish different contributions to the anomalous Hall effect. For example, under the assumption of a single type of scatterer, the anomalous Hall resistivity follows a function of the sum of a linear term from skew scattering \cite{SMIT1955877,SMIT195839} and a quadratic term from both the intrinsic and side jump mechanisms \cite{PhysRevB.2.4559}. However, the experimental measurements do not obey such a single-type scatterer relationship because multiple impurities (including phonons, magnons and interface/surface roughness) always exist that scatter the transport electrons \cite{PhysRevLett.103.087206,Hou_2012,PhysRevB.89.220406,doi:10.1143/JPSJ.81.083704,PhysRevB.83.125122,Grigoryan2016Scaling}. In addition to the above considerations, Hou \textit{et al.} \cite{PhysRevLett.114.217203} systemically studied the anomalous Hall effect in iron with changing thickness of the Fe film and temperature and reported a general scaling form of the anomalous Hall resistivity within the framework of multiple scattering \cite{PhysRevLett.114.217203,PhysRevB.85.220403}.

In this sense, our first-principles method can be used to study the extrinsic contributions by investigating the scaling law. 
And considering only static impurities, we have the following form of the scaling \cite{PhysRevLett.114.217203,PhysRevB.85.220403}:
\begin{eqnarray}
\sigma^{\mathrm{AH}}=-\alpha\sigma_{zz}-\beta_0
\label{sleq}
\end{eqnarray}
where $\alpha$ is the contribution from skew scattering and $\beta_0$ is the contribution from both the side jump and Berry curvature (intrinsic).

To confirm the above scaling law, we set up a clean structure of Fe and introduce few static impurities to calculate the anomalous Hall effect, e.g. Fe$_{1-c}$Pt$_c$ alloy at the condition of different $c$ as shown in Fig.~\ref{linear}. Here for a clear view of the data, we use a log-log scale to plot the calculated results of the longitudinal conductivity $\sigma_{zz}$ and the corresponding anomalous Hall conductivity $\sigma^{\mathrm{AH}}$, the red dash line represents the linear fitting when $c<0.01$. It can be seen that, the results agree well with the above Eq.~(\ref{sleq}) when $c<0.01$, which shows a good linear relationship between $\sigma_{zz}$ and $\sigma^{\mathrm{AH}}$. However, when $c>0.01$, the results exceed the scaling law of Eq.~(\ref{sleq}), this is because that, when $c$ becomes larger, the calculated anomalous Hall effect will be from the property of an alloy, which beyond the assumption of the clean bulk Fe with barely doped static impurities, and Pt will give more contribution to break the scaling law.

\begin{table*}[tbp]
	\begin{tabular}{C{3cm}C{3cm}C{3cm}C{3cm}C{3cm}}
		\hline\hline
		Impurity & $\alpha$ & $\alpha_{exp}$ & $\beta_0$ $[10^6\Omega^{-1}cm^{-1}]$ & $M_{orb}$ [$\mu_B$] \\ \hline
		C  &  0.0044 $\pm$ 0.0009  & --- & -0.00426 $\pm$ 0.00097 & -0.001 \\ 
		Cr & -0.00394 $\pm$ 0.00008 & -0.027 &  0.00149 $\pm$ 0.0002 &  0.02  \\ 
		Cu & -0.02391 $\pm$ 0.0002  & -0.025 &  0.01516 $\pm$ 0.0006 &  0.006  \\ 
		Pd & -0.01858 $\pm$ 0.00181 & --- &  0.01934 $\pm$ 0.01776 &  0.038  \\ 
		Ag & -0.02288 $\pm$ 0.0005 & --- &  0.0109 $\pm$ 0.00116 &  0.016  \\ 
		Pt & -0.0252 $\pm$ 0.001 & --- &  0.01909 $\pm$ 0.00205 &  0.037  \\
		\hline\hline
	\end{tabular}
	\caption{Fitting parameters $\alpha$ and $\beta_0$ from linear fitting of the anomalous Hall conductivity in Fig.~\ref{sl} for different kinds of impurities and their corresponding orbital momentum $M_{orb}$. Here the related values of $\alpha_{exp}$ from experimental measurements \cite{Shiomi2013} are shown for comparison.}
	\label{par}
\end{table*}

\begin{figure}[tbp]
	\includegraphics[width=\columnwidth]{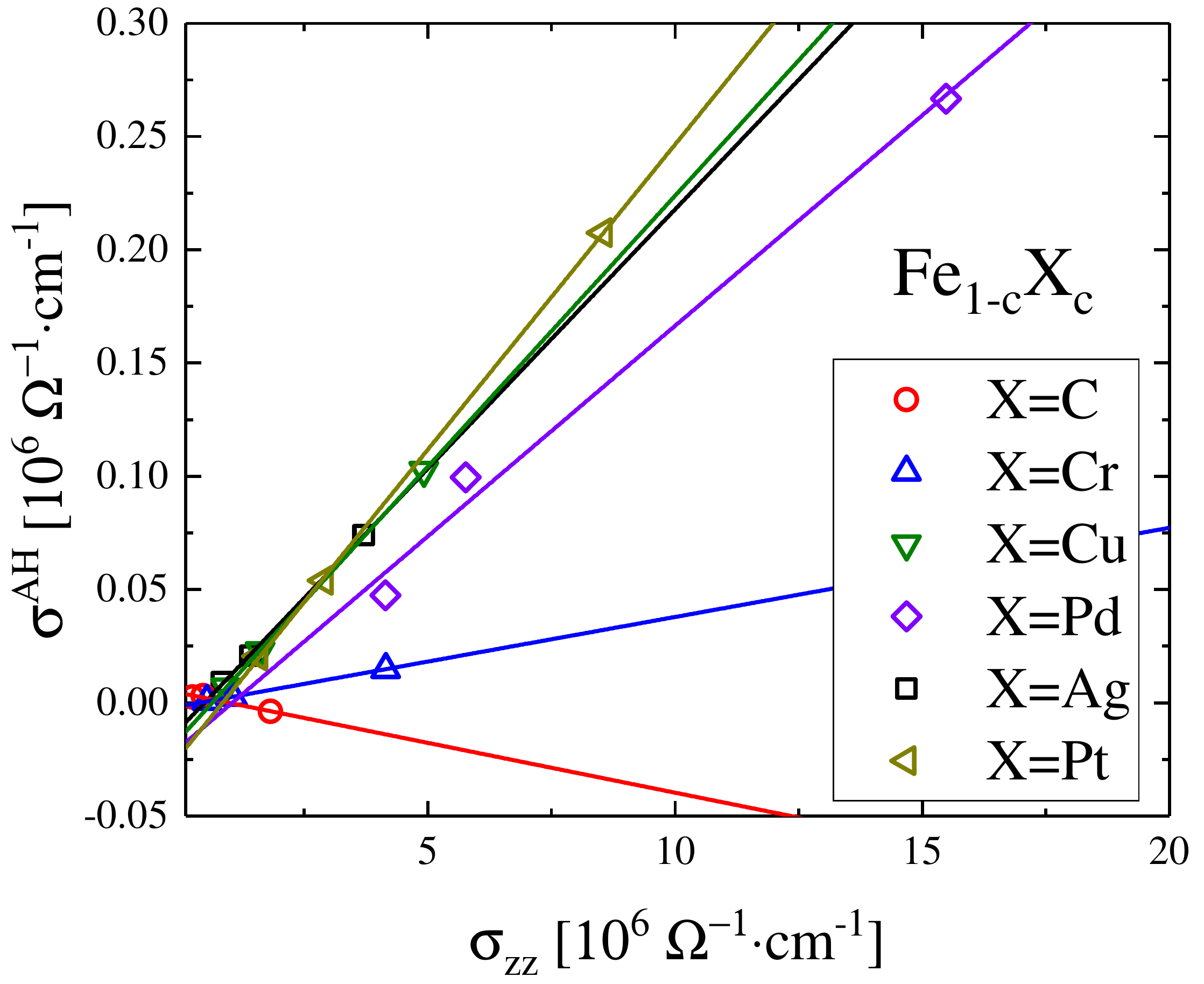}
	\caption{Anomalous Hall conductivity $\sigma^{\rm AH}=\Theta^{\rm AH}/\rho_{zz}$ of Fe$_{1-c}$X$_c$ as a function of the longitudinal conductivity $\sigma_{zz}=1/\rho_{zz}$, where X=C, Cr, Cu, Pd, Ag, Pt denotes the impurities, and $c\in\{0.001,0.004,0.007\}$ is the corresponding concentration.}
	\label{sl}
\end{figure}

On top of the above considerations, we only introduce sparse impurities ($<1\%$) of different kinds of elements (C, Cr, Cu, Pd, Ag and Pt) to investigate the scaling law of the anomalous Hall effect. And due to the sparse impurities, the contribution to the anomalous Hall effect should entirely come from impurity scattering under the same Fe background. The calculated anomalous Hall conductivity $\sigma^{\mathrm{AH}}$ as a function of the longitudinal conductivity $\sigma_{zz}$ is plotted in Fig.~\ref{sl}. And we find good linear relationship between $\sigma^{\mathrm{AH}}$ and $\sigma_{zz}$ as shown in Fig.~\ref{sl} for all kinds of impurities. Thus, we can fit the corresponding parameters $\alpha$ and $\beta_0$, respectively. The fitting parameter values are shown in Tab.~\ref{par}, together with the values from experiment \cite{Shiomi2013} for comparison. It can be seen that our result with Cu impurity shows good agreement with the experiment, but one magnitude smaller with Cr impurity. This difference with Cr impurity may come from the complex antiferromagnetic order of Cr in Fe in the experiments, which could introduce more contributions to the anomalous Hall effect and needs further investigations.

For the skew scattering coefficient, all the impurities except for the C impurity have a negative $\alpha$. Conversely, the $\beta_0$ of the C impurity is negative, while the other impurities have positive values. This sign change between skew scattering and side jump can be understood by the direction of the orbital momentum $M_{orb}$, as the anomalous Hall conductivity of ferromagnetic systems has been proven to be fully determined by the response of the orbital momentum \cite{PhysRevB.88.134422}, where the Hall current is proportional to the orbital momentum. Thus, we also calculate the corresponding orbital momentum of the impurities under the Fe background, and the results are also shown in Tab.~\ref{par}. The sign of the orbital momentum of C is different from that of the other impurities. In this sense, the anomalous Hall conductivity with C impurities will have a different sign, leading to the different signs of $\alpha$ and $\beta_0$.

The intrinsic anomalous Hall conductivity from Berry curvature calculations is approximately $\sigma^{\mathrm{AH}}_{intr}=751~\Omega^{-1}cm^{-1}$  \cite{PhysRevLett.92.037204,PhysRevB.76.195109} at zero temperature. Therefore, as the longitudinal conductivity $\sigma_{zz}$ is on the order of $10^6\Omega^{-1}cm^{-1}$, as shown in Fig.~\ref{sl}, the corresponding anomalous Hall conductivity from skew scattering $\sigma^{\mathrm{AH}}_{ss}=\alpha\sigma_{zz}$ will be on the same order as that from the side jump $\sigma^{\mathrm{AH}}_{sj}=\beta_0-\sigma^{\mathrm{AH}}_{intr}$ but much larger than the intrinsic $\sigma^{\mathrm{AH}}_{intr}$ for bcc Fe with sparse impurities.

\subsection{Temperature dependence}
\begin{figure}[tbp]
	\includegraphics[width=\columnwidth]{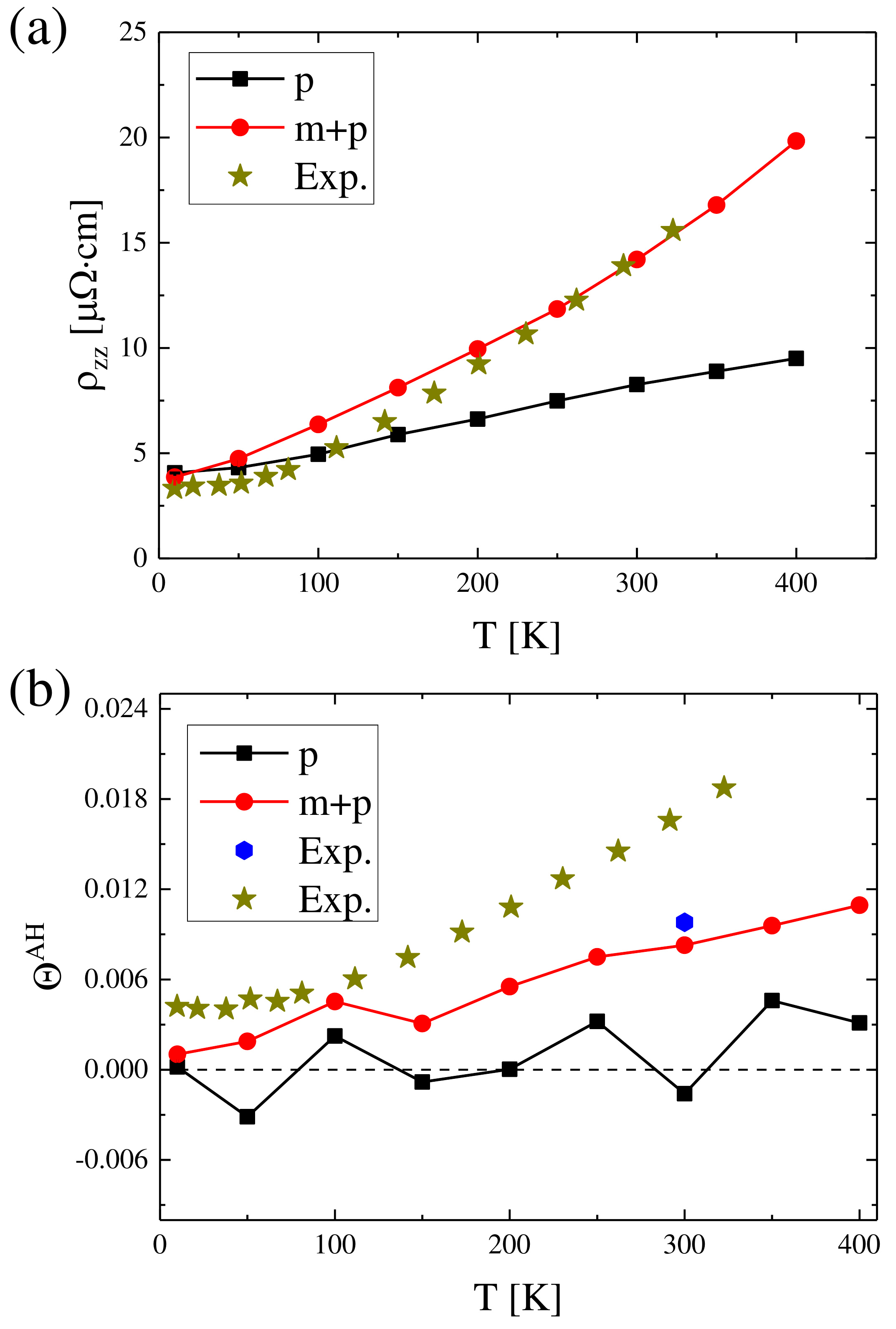}
	\caption{(a) Longitudinal resistivity $\rho_{zz}$ as a function of temperature. The black squares and the red dots are calculated without and with magnons. The stars show the experimental data measured in a thin film with a 33 nm thickness~\cite{PhysRevLett.114.217203}. (b) Corresponding anomalous Hall angle $\Theta^{\rm AH}$ as a function of temperature. The stars~\cite{PhysRevLett.114.217203} and blue hexagon~\cite{PhysRev.156.637} are from experimental measurements and plotted for comparison.}
	\label{temp}
\end{figure}

We first address the longitudinal resistivity ($\rho_{zz}$) of Fe before discussing the anomalous Hall effect. The calculated results are plotted in Fig.~\ref{temp} (a), where ``p'' denotes calculation with phonons only, while ``m+p'' represents calculation with both magnons and phonons. For comparison, we also plot the experimental data measured in a high quality 33 nm thick thin film \cite{PhysRevLett.114.217203}. The $\rho_{zz}$ calculated with only phonons is far from the experimental results, while when both phonons and magnons are considered, the calculated $\rho_{zz}$ agrees well with the experimental data in a broadened temperature zone. These results indicate that even though the fluctuation of the magnetization is very small (e.g., $\langle\cos\vartheta_i\rangle\simeq0.97$ for T=300 K), the magnons contribute to electronic transport significantly. Thus, the contribution from magnons to the anomalous Hall effect should also be studied accordingly.

The corresponding anomalous Hall angle $\Theta^{\rm AH}$ as a function of temperature is plotted in Fig.~\ref{temp} (b) together with the data (stars and blue hexagon) from the experiments \cite{PhysRevLett.114.217203,PhysRev.156.637}. We can see that, the $\Theta^{\rm AH}$ calculated with both magnons and phonons is close to one experimental result (blue hexagon) at room temperature but much smaller than the other (stars). This is because that, the measurements in Ref.~\cite{PhysRevLett.114.217203} were applied on a 33 nm thin film, in which, we believe that, there should be an extra strong contribution from the surface roughness as reported in Ref.~\cite{Grigoryan2016Scaling}. Moreover, we obtain a fluctuation in the calculation with only phonons, which has never been previously reported in the literature, to the best of our knowledge. This fluctuation almost disappears when including magnons, with only a small peak remaining at approximately T=100 K, which is why a general measurement in experiments cannot reveal this fluctuation effect. However, if one could suppress magnon excitation by applying a strong external magnetic field, then this phenomenon could be measured in the future.

The extrinsic contributions to the anomalous Hall effect are proportional to $\rho_{zz}$ or $\rho_{zz}^2$ \cite{PhysRevLett.114.217203,PhysRevB.85.220403}, and $\rho_{zz}$ was proven to increase monotonically with temperature, as shown in Fig.~\ref{temp} (a); 
therefore, the fluctuation of $\Theta^{\rm AH}$ with only phonons cannot come from the extrinsic part of the anomalous Hall effect. The intrinsic part of the anomalous Hall effect comes from the Berry curvature \cite{PhysRevLett.92.037204,PhysRevLett.93.206602,PhysRevB.76.195109,PhysRevB.74.195118}, which is dominated by the detailed band structure around the Fermi energy \cite{PhysRevLett.93.206602,PhysRevB.76.195109}. In addition to the above information, we realized that the Berry curvature from each $\mathbf{k}$ point inside the Brillouin zone can contribute not only positively but also negatively to the anomalous Hall effect and the anomalous Hall conductivity is dominated by few hot spots in the Brillouin zone, as reported in Ref.~\cite{PhysRevLett.92.037204,PhysRevB.74.195118}. Moreover, the atomic displacements arising from phonons directly affect the detailed band structure, which gives us strong confidence that the fluctuation of $\Theta^{\rm AH}$ comes from the changing Berry curvature, which will be confirmed in Sec.~\ref{os} with individual Berry curvature calculations.

\begin{figure}[tbp]
	\includegraphics[width=\columnwidth]{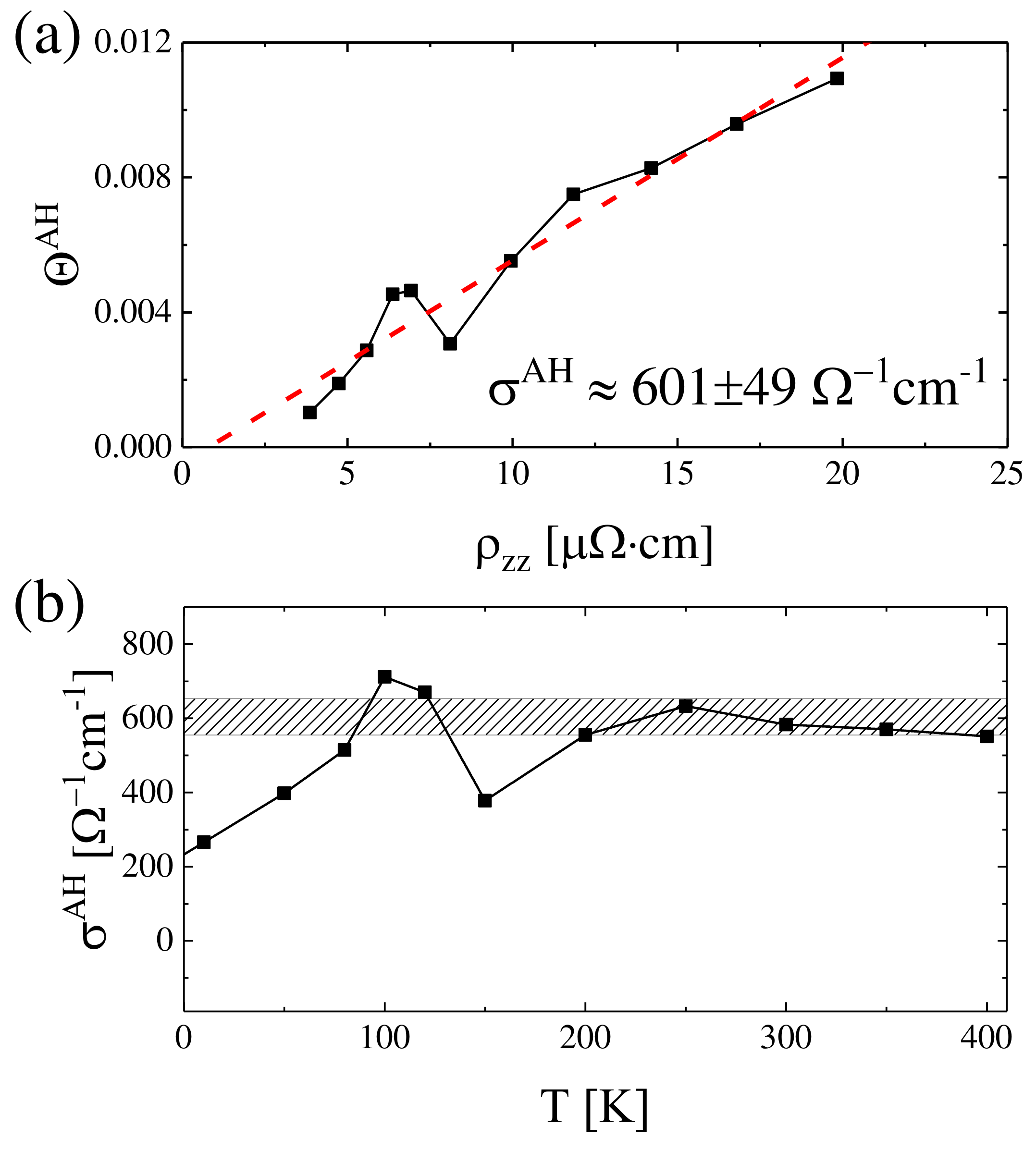}
	\caption{(a) Anomalous Hall angle as a function of $\rho_{zz}$ from our calculations with both magnons and phonons and a linear fitting to approach the temperature-independent anomalous Hall conductivity. (b) Corresponding anomalous Hall conductivity obtained by $\sigma^{\rm AH}=\Theta^{\rm AH}/\rho_{zz}$ as a function of temperature, where the shadow zone represents the results from the upper panel.}
	\label{sigma}
\end{figure}

On top of the above discussions, the results in Fig.~\ref{temp} can be reinterpreted in a way that: the whole temperature dependent anomalous Hall effect includes both the extrinsic contributions from phonons and magnons (proportional to $\rho_{zz}$ or $\rho_{zz}^2$) and the intrinsic contribution. Therefore, the competition between the monotonic extrinsic effect and fluctuated intrinsic effect will determine the shape of the curves in Fig.~\ref{temp} (b). Therefore, the calculated results in Fig.~\ref{temp} (b) indicate that, when introduce phonons only, the fluctuated intrinsic contribution dominates the anomalous Hall effect and then the anomalous Hall angle fluctuates with increasing temperature. When include both phonon and magnons, the monotonic extrinsic contributions dominate, and then the anomalous Hall angle increases with increasing temperature and only a small peak around T=100 K left. However, these assumptions should be verified by further investigations in the future, which needs an emergent method to separate all the contributions under one single frame.

The anomalous Hall conductivity $\sigma^{\rm AH}$ is also studied with both phonons and magnons contributions. This conductivity is related to the longitudinal resistivity and anomalous Hall angle by $\Theta^{\rm AH}=\sigma^{\rm AH}\rho_{zz}$. Thus, if we assume that the anomalous Hall conductivity is independent of the temperature, then $\sigma^{\rm AH}$ can be obtained from the slope of the function $\Theta^{\rm AH}(\rho_{zz})$, which is shown in Fig.~\ref{sigma} (a). 
Except for the fluctuation around T=100 K induced by the corresponding peak in Fig.~\ref{temp} (b), $\Theta^{\rm AH}$ and $\rho_{zz}$ have an approximately linear relation, and one can obtain the anomalous Hall conductivity as $\sigma^{\rm AH}\simeq601\pm49~\Omega^{-1}cm^{-1}$. Furthermore, as shown in Fig.~\ref{sigma} (b), we calculate the anomalous Hall conductivity point by point using $\sigma^{\rm AH}=\Theta^{\rm AH}/\rho_{zz}$, and the shadow zone represents the results from Fig.~\ref{sigma} (a) for comparison. Together with the analysis from Fig.~\ref{sigma} (a), we can conclude that similar to the $\Theta^{\rm AH}$ in Fig.~\ref{temp} (b) with both phonons and magnons contributions, the anomalous Hall conductivity fluctuates at low temperature and becomes constant when the temperature is sufficiently high.

\subsection{Berry curvature with phonon}
\label{os}
To circumstantially verify the above assumption on the nature of the fluctuation of the anomalous Hall effect with only phonon, we set up a 3$\times$3$\times$3 supercell of the bcc Fe (54 atoms) and introduce random atomic displacements which is the same with that in the transport calculation in Fig.~\ref{temp} at T=10 K with only phonon, then the Berry curvature can be studied with including the phonon contribution. And as our transport code is not able to individually calculate the Berry curvature for now (all contributions are entangled together), the results in this sub-section are calculated using third-party codes.

Technically, we carry out the calculations of the electronic structure using the VASP (Vienna ab-initio simulation package) code \cite{PhysRevB.47.558,PhysRevB.54.11169}, and all the calculations in this section are based on DFT and the generalized gradient approximation (GGA) with an interpolation formula according to Vosko, Wilk, and Nusair \cite{1980CaJPh} and a plane-wave basis set within the framework of the projector augmented wave (PAW) method \cite{PhysRevB.50.17953,PhysRevB.59.1758}. The cut-off energy for the basis is 500 eV, and the convergence criterion for the electron density self-consistency cycles is $10^{-5}$ eV for the whole supercell. In the Brillouin zone, we sample ($3\times3\times3$) k-point grids using the Monk-horst-Pack scheme \cite{PhysRevB.13.5188} to make sure the results converged. Also for convenient of the further study of the anomalous Hall effect, the spin-orbit coupling is introduced.

\begin{figure}[tbp]
	\includegraphics[width=\columnwidth]{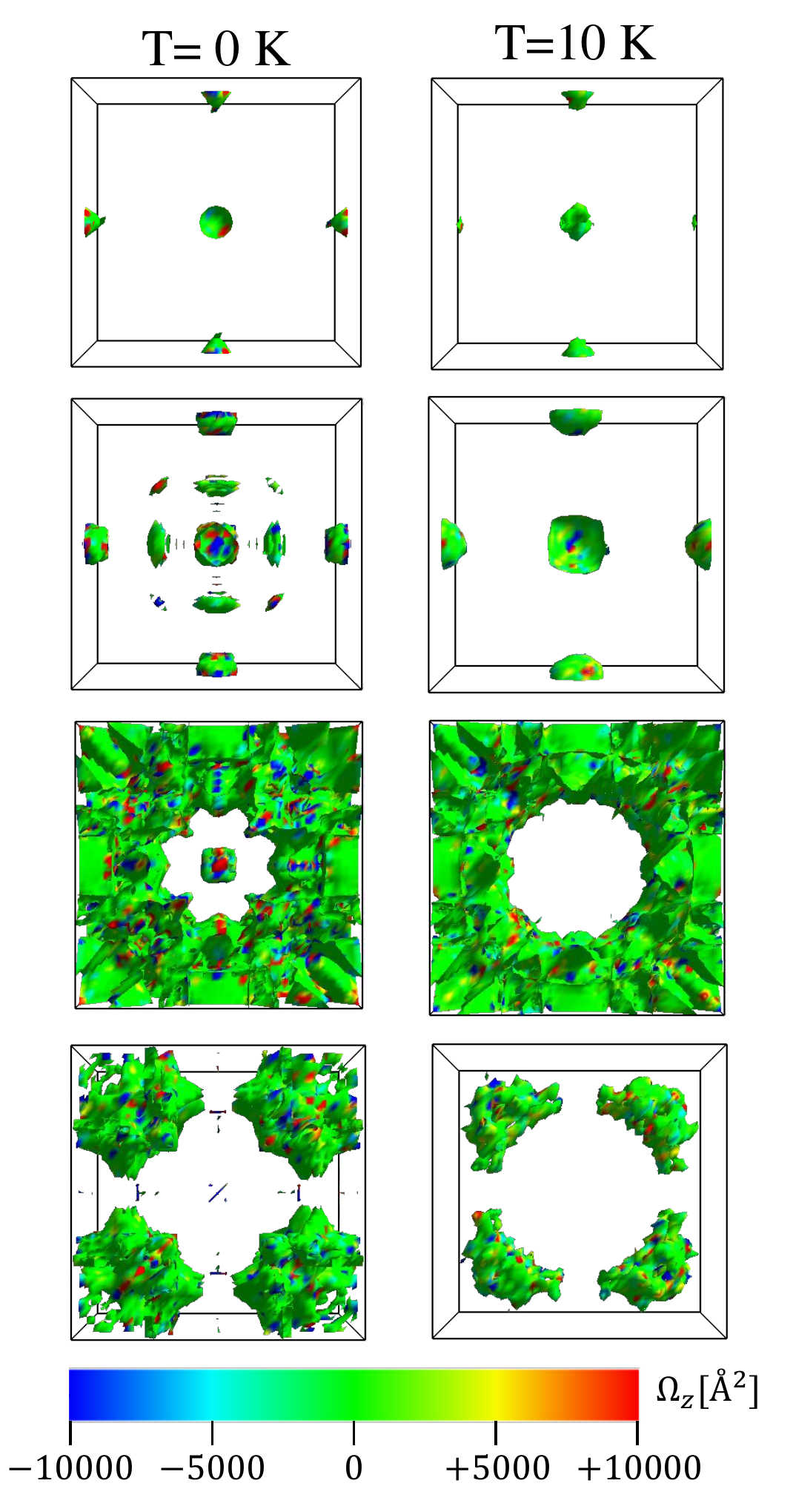}
	\caption{The Berry curvature of four typical bands that cross over the Fermi energy at T=0 K and T=10 K respectively, where the contour profiles stand for the corresponding Fermi surface of each bands and the color represents the value of the Berry curvature according to the following color bar.}
	\label{e-ahc}
\end{figure}

For the calculation of the anomalous Hall conductivity from the contribution of the Berry curvature, a well-known formula \cite{PhysRevLett.92.037204,PhysRevB.76.195109} will be used, and when the magnetization parallel to $z$ direction, reads,
\begin{eqnarray}
	\sigma^{\rm AH}_{intr}=-\frac{e^2}{\hbar}\int_{BZ}\frac{d^3k}{(2\pi)^3}\Omega^z(\mathbf{k})
\end{eqnarray}
where $\hbar$ is the Planck constant, ``$BZ$'' represent the integration over the total Brillouin zone, $k$ is the wave vector, and $\Omega^z(\mathbf{k})$ is the sum of the Berry curvatures over the occupied bands for each $\mathbf{k}$:
\begin{eqnarray}
	\Omega^z(\mathbf{k})=\sum_n f_n \Omega_n^z(\mathbf{k})
\end{eqnarray}
with the band number $n$ and the corresponding equilibrium Fermi-Dirac distribution function $f_n$, and the Berry curvature arises from the Kubo-formula derivation \cite{PhysRevLett.49.405}:
\begin{eqnarray}
	\Omega_n^z(\mathbf{k})=-\sum_{n'\neq n}\frac{2\mathrm{Im}\langle\psi_{n\mathbf{k}}\vert v_x \vert\psi_{n'\mathbf{k}}\rangle \langle\psi_{n'\mathbf{k}}\vert v_y \vert\psi_{n\mathbf{k}}\rangle}{(\omega_{n'}-\omega_n)^2}
\end{eqnarray}
where the energy of each band $E_n=\hbar\omega_n$, $v_{x,y}$ are velocity operators and $\psi$ is the wave function.

The above formula had already been generated in the open source code ``WANNIER90'' \cite{Pizzi_2020} and ``Wannier Berri'' \cite{wannier-berri1,wannier-berri2} with the maximally localized generalized Wannier functions (MLWFs) \cite{PhysRevB.56.12847,PhysRevB.65.035109,RevModPhys.84.1419} which can connected to our previous VASP results conveniently. In addition, the direction of the magnetization is parallel to the bcc (001) direction, which is the $z$-axis in our global coordinate system. And we use a three-dimensional $\mathbf{k}$ mesh in the total Brillouin zone with the spacing of $\mathbf{k}$-points being $\Delta\mathbf{k}\simeq\frac{2\pi}{\textit{Len}}$, where $\textit{Len}=300$,  typically. Moreover, to make the calculation more precise around the typical $\mathbf{k}$ points with major contribution to the Berry curvature, the adaptive recursive refinement algorithm \cite{wannier-berri1} is used, and we calculate 100 iterations to make sure the Berry curvature calculations converged. 

The calculated intrinsic anomalous Hall conductivities from Berry curvature are about $\sigma^{\rm AH}_{intr}\simeq666~\Omega^{-1}cm^{-1}$ at T=0 K and $\sigma^{\rm AH}_{intr}\simeq388~\Omega^{-1}cm^{-1}$ for one configuration at T=10 K, respectively, which reveals that the Berry curvature is quite sensitive to the phonons. Moreover, for clear view of the detail Berry curvature distortion, we plot the corresponding Berry curvature of four typical bands that cross over the Fermi energy in Fig.~\ref{e-ahc}, where the contour profiles stand for the corresponding Fermi surface of each bands and the color represents the value of the Berry curvature according to the following color bar. It can be seen that, after introducing the phonon at T=10 K, both the blue zone and red zone are smoothed down, means that the hot spots of the Berry curvature are suppressed by the phonon. This can be understood in a way that, the Berry curvature gives a large contribution when the Fermi surface lies in a spin-orbit induced gap \cite{PhysRevLett.92.037204}, and by comparing the contour profiles (Fermi surface) at T=10 K with that at T=0 K as shown in Fig.~\ref{e-ahc}, we can conclude that the corresponding bands that cross over the Fermi energy can be easily twisted by the phonon, which end up with the changing of the Berry curvature accordingly. 

\section{Conclusion}
In this work, we developed a first-principles method based on EMTOs and studied the anomalous Hall effect in ferromagnetic metals. We systematically investigated the scaling law of the anomalous Hall effect by introducing sparse impurities and studied the temperature effect by taking into account both phonons and magnons. We successfully separated the skew scattering contribution to the anomalous Hall effect from the side jump and intrinsic contributions. The numerical results showed that the contributions from skew scattering and side jump are on the same order, and they are both much larger than the intrinsic mechanics in the doped system.

Moreover, with the study of the contributions from magnons and phonons separately, we found that the magnons have a significant contribution to the anomalous Hall effect. Specifically, we predicted a remarkable fluctuation behavior of the anomalous Hall angle when considering only phonons, which needs to be checked by experiments in the future.

\begin{acknowledgments}
The authors are grateful to Levente Vitos for providing the bulk EMTO-CPA self-consistent code and Youqi Ke for helpful discussion on the two-electrode EMTOs method with the fully relativistic effect, and this work was supported by the National Key Research and Development Program of China (grant Nos. 2018YFB0407600, 2017YFA0206202 and 2016YFA0300702), National Natural Science Foundation of China (grant No. 11804266) and Shaanxi Province Science and Technology Innovation Project (grant 2019TSLGY08-04). K. X. is supported by the National Key Research and Development Program of China (grant Nos. 2017YFA0303300 and 2018YFB0407601), the National Natural Science Foundation of China (grant Nos. 61774017, 11734004 and 21421003), and NSAF (grant No. U1930402).
\end{acknowledgments}


\begin{thebibliography}{78}%
	\makeatletter
	\providecommand \@ifxundefined [1]{%
		\@ifx{#1\undefined}
	}%
	\providecommand \@ifnum [1]{%
		\ifnum #1\expandafter \@firstoftwo
		\else \expandafter \@secondoftwo
		\fi
	}%
	\providecommand \@ifx [1]{%
		\ifx #1\expandafter \@firstoftwo
		\else \expandafter \@secondoftwo
		\fi
	}%
	\providecommand \natexlab [1]{#1}%
	\providecommand \enquote  [1]{``#1''}%
	\providecommand \bibnamefont  [1]{#1}%
	\providecommand \bibfnamefont [1]{#1}%
	\providecommand \citenamefont [1]{#1}%
	\providecommand \href@noop [0]{\@secondoftwo}%
	\providecommand \href [0]{\begingroup \@sanitize@url \@href}%
	\providecommand \@href[1]{\@@startlink{#1}\@@href}%
	\providecommand \@@href[1]{\endgroup#1\@@endlink}%
	\providecommand \@sanitize@url [0]{\catcode `\\12\catcode `\$12\catcode
		`\&12\catcode `\#12\catcode `\^12\catcode `\_12\catcode `\%12\relax}%
	\providecommand \@@startlink[1]{}%
	\providecommand \@@endlink[0]{}%
	\providecommand \url  [0]{\begingroup\@sanitize@url \@url }%
	\providecommand \@url [1]{\endgroup\@href {#1}{\urlprefix }}%
	\providecommand \urlprefix  [0]{URL }%
	\providecommand \Eprint [0]{\href }%
	\providecommand \doibase [0]{https://doi.org/}%
	\providecommand \selectlanguage [0]{\@gobble}%
	\providecommand \bibinfo  [0]{\@secondoftwo}%
	\providecommand \bibfield  [0]{\@secondoftwo}%
	\providecommand \translation [1]{[#1]}%
	\providecommand \BibitemOpen [0]{}%
	\providecommand \bibitemStop [0]{}%
	\providecommand \bibitemNoStop [0]{.\EOS\space}%
	\providecommand \EOS [0]{\spacefactor3000\relax}%
	\providecommand \BibitemShut  [1]{\csname bibitem#1\endcsname}%
	\let\auto@bib@innerbib\@empty
	\bibitem [{\citenamefont {{Hall}}(1880)}]{Hall1880}%
	\BibitemOpen
	\bibfield  {author} {\bibinfo {author} {\bibfnamefont {E.~H.}\ \bibnamefont
			{{Hall}}},\ }\href {https://doi.org/10.1088/1478-7814/4/1/335} {\bibfield
		{journal} {\bibinfo  {journal} {Proceedings of the Physical Society of
				London}\ }\textbf {\bibinfo {volume} {4}},\ \bibinfo {pages} {325} (\bibinfo
		{year} {1880})}\BibitemShut {NoStop}%
	\bibitem [{\citenamefont {Karplus}\ and\ \citenamefont
		{Luttinger}(1954)}]{PhysRev.95.1154}%
	\BibitemOpen
	\bibfield  {author} {\bibinfo {author} {\bibfnamefont {R.}~\bibnamefont
			{Karplus}}\ and\ \bibinfo {author} {\bibfnamefont {J.~M.}\ \bibnamefont
			{Luttinger}},\ }\href {https://doi.org/10.1103/PhysRev.95.1154} {\bibfield
		{journal} {\bibinfo  {journal} {Phys. Rev.}\ }\textbf {\bibinfo {volume}
			{95}},\ \bibinfo {pages} {1154} (\bibinfo {year} {1954})}\BibitemShut
	{NoStop}%
	\bibitem [{\citenamefont {Jungwirth}\ \emph {et~al.}(2002)\citenamefont
		{Jungwirth}, \citenamefont {Niu},\ and\ \citenamefont
		{MacDonald}}]{PhysRevLett.88.207208}%
	\BibitemOpen
	\bibfield  {author} {\bibinfo {author} {\bibfnamefont {T.}~\bibnamefont
			{Jungwirth}}, \bibinfo {author} {\bibfnamefont {Q.}~\bibnamefont {Niu}},\
		and\ \bibinfo {author} {\bibfnamefont {A.~H.}\ \bibnamefont {MacDonald}},\
	}\href {https://doi.org/10.1103/PhysRevLett.88.207208} {\bibfield  {journal}
		{\bibinfo  {journal} {Phys. Rev. Lett.}\ }\textbf {\bibinfo {volume} {88}},\
		\bibinfo {pages} {207208} (\bibinfo {year} {2002})}\BibitemShut {NoStop}%
	\bibitem [{\citenamefont {Nagaosa}\ \emph {et~al.}(2010)\citenamefont
		{Nagaosa}, \citenamefont {Sinova}, \citenamefont {Onoda}, \citenamefont
		{MacDonald},\ and\ \citenamefont {Ong}}]{RevModPhys.82.1539}%
	\BibitemOpen
	\bibfield  {author} {\bibinfo {author} {\bibfnamefont {N.}~\bibnamefont
			{Nagaosa}}, \bibinfo {author} {\bibfnamefont {J.}~\bibnamefont {Sinova}},
		\bibinfo {author} {\bibfnamefont {S.}~\bibnamefont {Onoda}}, \bibinfo
		{author} {\bibfnamefont {A.~H.}\ \bibnamefont {MacDonald}},\ and\ \bibinfo
		{author} {\bibfnamefont {N.~P.}\ \bibnamefont {Ong}},\ }\href
	{https://doi.org/10.1103/RevModPhys.82.1539} {\bibfield  {journal} {\bibinfo
			{journal} {Rev. Mod. Phys.}\ }\textbf {\bibinfo {volume} {82}},\ \bibinfo
		{pages} {1539} (\bibinfo {year} {2010})}\BibitemShut {NoStop}%
	\bibitem [{\citenamefont {Kondo}(1962)}]{Kondo01041962}%
	\BibitemOpen
	\bibfield  {author} {\bibinfo {author} {\bibfnamefont {J.}~\bibnamefont
			{Kondo}},\ }\href {https://doi.org/10.1143/PTP.27.772} {\bibfield  {journal}
		{\bibinfo  {journal} {Progress of Theoretical Physics}\ }\textbf {\bibinfo
			{volume} {27}},\ \bibinfo {pages} {772} (\bibinfo {year} {1962})}\BibitemShut
	{NoStop}%
	\bibitem [{\citenamefont {Toyosaki}\ \emph {et~al.}(2004)\citenamefont
		{Toyosaki}, \citenamefont {Fukumura}, \citenamefont {Yamada}, \citenamefont
		{Nakajima}, \citenamefont {Chikyow}, \citenamefont {Hasegawa}, \citenamefont
		{Koinuma},\ and\ \citenamefont {Kawasaki}}]{natm3.4}%
	\BibitemOpen
	\bibfield  {author} {\bibinfo {author} {\bibfnamefont {H.}~\bibnamefont
			{Toyosaki}}, \bibinfo {author} {\bibfnamefont {T.}~\bibnamefont {Fukumura}},
		\bibinfo {author} {\bibfnamefont {Y.}~\bibnamefont {Yamada}}, \bibinfo
		{author} {\bibfnamefont {K.}~\bibnamefont {Nakajima}}, \bibinfo {author}
		{\bibfnamefont {T.}~\bibnamefont {Chikyow}}, \bibinfo {author} {\bibfnamefont
			{T.}~\bibnamefont {Hasegawa}}, \bibinfo {author} {\bibfnamefont
			{H.}~\bibnamefont {Koinuma}},\ and\ \bibinfo {author} {\bibfnamefont
			{M.}~\bibnamefont {Kawasaki}},\ }\href
	{https://doi.org/http://www.nature.com/nmat/journal/v3/n4/suppinfo/nmat1099_S1.html}
	{\bibfield  {journal} {\bibinfo  {journal} {Nat Mater}\ }\textbf {\bibinfo
			{volume} {3}},\ \bibinfo {pages} {221} (\bibinfo {year} {2004})}\BibitemShut
	{NoStop}%
	\bibitem [{\citenamefont {Shinde}\ \emph {et~al.}(2004)\citenamefont {Shinde},
		\citenamefont {Ogale}, \citenamefont {Higgins}, \citenamefont {Zheng},
		\citenamefont {Millis}, \citenamefont {Kulkarni}, \citenamefont {Ramesh},
		\citenamefont {Greene},\ and\ \citenamefont
		{Venkatesan}}]{PhysRevLett.92.166601}%
	\BibitemOpen
	\bibfield  {author} {\bibinfo {author} {\bibfnamefont {S.~R.}\ \bibnamefont
			{Shinde}}, \bibinfo {author} {\bibfnamefont {S.~B.}\ \bibnamefont {Ogale}},
		\bibinfo {author} {\bibfnamefont {J.~S.}\ \bibnamefont {Higgins}}, \bibinfo
		{author} {\bibfnamefont {H.}~\bibnamefont {Zheng}}, \bibinfo {author}
		{\bibfnamefont {A.~J.}\ \bibnamefont {Millis}}, \bibinfo {author}
		{\bibfnamefont {V.~N.}\ \bibnamefont {Kulkarni}}, \bibinfo {author}
		{\bibfnamefont {R.}~\bibnamefont {Ramesh}}, \bibinfo {author} {\bibfnamefont
			{R.~L.}\ \bibnamefont {Greene}},\ and\ \bibinfo {author} {\bibfnamefont
			{T.}~\bibnamefont {Venkatesan}},\ }\href
	{https://doi.org/10.1103/PhysRevLett.92.166601} {\bibfield  {journal}
		{\bibinfo  {journal} {Phys. Rev. Lett.}\ }\textbf {\bibinfo {volume} {92}},\
		\bibinfo {pages} {166601} (\bibinfo {year} {2004})}\BibitemShut {NoStop}%
	\bibitem [{\citenamefont {Fang}\ \emph {et~al.}(2003)\citenamefont {Fang},
		\citenamefont {Nagaosa}, \citenamefont {Takahashi}, \citenamefont {Asamitsu},
		\citenamefont {Mathieu}, \citenamefont {Ogasawara}, \citenamefont {Yamada},
		\citenamefont {Kawasaki}, \citenamefont {Tokura},\ and\ \citenamefont
		{Terakura}}]{Fang92}%
	\BibitemOpen
	\bibfield  {author} {\bibinfo {author} {\bibfnamefont {Z.}~\bibnamefont
			{Fang}}, \bibinfo {author} {\bibfnamefont {N.}~\bibnamefont {Nagaosa}},
		\bibinfo {author} {\bibfnamefont {K.~S.}\ \bibnamefont {Takahashi}}, \bibinfo
		{author} {\bibfnamefont {A.}~\bibnamefont {Asamitsu}}, \bibinfo {author}
		{\bibfnamefont {R.}~\bibnamefont {Mathieu}}, \bibinfo {author} {\bibfnamefont
			{T.}~\bibnamefont {Ogasawara}}, \bibinfo {author} {\bibfnamefont
			{H.}~\bibnamefont {Yamada}}, \bibinfo {author} {\bibfnamefont
			{M.}~\bibnamefont {Kawasaki}}, \bibinfo {author} {\bibfnamefont
			{Y.}~\bibnamefont {Tokura}},\ and\ \bibinfo {author} {\bibfnamefont
			{K.}~\bibnamefont {Terakura}},\ }\href
	{https://doi.org/10.1126/science.1089408} {\bibfield  {journal} {\bibinfo
			{journal} {Science}\ }\textbf {\bibinfo {volume} {302}},\ \bibinfo {pages}
		{92} (\bibinfo {year} {2003})}\BibitemShut {NoStop}%
	\bibitem [{\citenamefont {Yao}\ \emph {et~al.}(2004)\citenamefont {Yao},
		\citenamefont {Kleinman}, \citenamefont {MacDonald}, \citenamefont {Sinova},
		\citenamefont {Jungwirth}, \citenamefont {Wang}, \citenamefont {Wang},\ and\
		\citenamefont {Niu}}]{PhysRevLett.92.037204}%
	\BibitemOpen
	\bibfield  {author} {\bibinfo {author} {\bibfnamefont {Y.}~\bibnamefont
			{Yao}}, \bibinfo {author} {\bibfnamefont {L.}~\bibnamefont {Kleinman}},
		\bibinfo {author} {\bibfnamefont {A.~H.}\ \bibnamefont {MacDonald}}, \bibinfo
		{author} {\bibfnamefont {J.}~\bibnamefont {Sinova}}, \bibinfo {author}
		{\bibfnamefont {T.}~\bibnamefont {Jungwirth}}, \bibinfo {author}
		{\bibfnamefont {D.-S.}\ \bibnamefont {Wang}}, \bibinfo {author}
		{\bibfnamefont {E.}~\bibnamefont {Wang}},\ and\ \bibinfo {author}
		{\bibfnamefont {Q.}~\bibnamefont {Niu}},\ }\href
	{https://doi.org/10.1103/PhysRevLett.92.037204} {\bibfield  {journal}
		{\bibinfo  {journal} {Phys. Rev. Lett.}\ }\textbf {\bibinfo {volume} {92}},\
		\bibinfo {pages} {037204} (\bibinfo {year} {2004})}\BibitemShut {NoStop}%
	\bibitem [{\citenamefont {Chang}\ and\ \citenamefont
		{Niu}(1996)}]{PhysRevB.53.7010}%
	\BibitemOpen
	\bibfield  {author} {\bibinfo {author} {\bibfnamefont {M.-C.}\ \bibnamefont
			{Chang}}\ and\ \bibinfo {author} {\bibfnamefont {Q.}~\bibnamefont {Niu}},\
	}\href {https://doi.org/10.1103/PhysRevB.53.7010} {\bibfield  {journal}
		{\bibinfo  {journal} {Phys. Rev. B}\ }\textbf {\bibinfo {volume} {53}},\
		\bibinfo {pages} {7010} (\bibinfo {year} {1996})}\BibitemShut {NoStop}%
	\bibitem [{\citenamefont {Sundaram}\ and\ \citenamefont
		{Niu}(1999)}]{PhysRevB.59.14915}%
	\BibitemOpen
	\bibfield  {author} {\bibinfo {author} {\bibfnamefont {G.}~\bibnamefont
			{Sundaram}}\ and\ \bibinfo {author} {\bibfnamefont {Q.}~\bibnamefont {Niu}},\
	}\href {https://doi.org/10.1103/PhysRevB.59.14915} {\bibfield  {journal}
		{\bibinfo  {journal} {Phys. Rev. B}\ }\textbf {\bibinfo {volume} {59}},\
		\bibinfo {pages} {14915} (\bibinfo {year} {1999})}\BibitemShut {NoStop}%
	\bibitem [{\citenamefont {Smit}(1955)}]{SMIT1955877}%
	\BibitemOpen
	\bibfield  {author} {\bibinfo {author} {\bibfnamefont {J.}~\bibnamefont
			{Smit}},\ }\href
	{https://doi.org/https://doi.org/10.1016/S0031-8914(55)92596-9} {\bibfield
		{journal} {\bibinfo  {journal} {Physica}\ }\textbf {\bibinfo {volume} {21}},\
		\bibinfo {pages} {877 } (\bibinfo {year} {1955})}\BibitemShut {NoStop}%
	\bibitem [{\citenamefont {Smit}(1958)}]{SMIT195839}%
	\BibitemOpen
	\bibfield  {author} {\bibinfo {author} {\bibfnamefont {J.}~\bibnamefont
			{Smit}},\ }\href
	{https://doi.org/https://doi.org/10.1016/S0031-8914(58)93541-9} {\bibfield
		{journal} {\bibinfo  {journal} {Physica}\ }\textbf {\bibinfo {volume} {24}},\
		\bibinfo {pages} {39 } (\bibinfo {year} {1958})}\BibitemShut {NoStop}%
	\bibitem [{\citenamefont {Berger}(1970)}]{PhysRevB.2.4559}%
	\BibitemOpen
	\bibfield  {author} {\bibinfo {author} {\bibfnamefont {L.}~\bibnamefont
			{Berger}},\ }\href {https://doi.org/10.1103/PhysRevB.2.4559} {\bibfield
		{journal} {\bibinfo  {journal} {Phys. Rev. B}\ }\textbf {\bibinfo {volume}
			{2}},\ \bibinfo {pages} {4559} (\bibinfo {year} {1970})}\BibitemShut
	{NoStop}%
	\bibitem [{\citenamefont {Haldane}(2004)}]{PhysRevLett.93.206602}%
	\BibitemOpen
	\bibfield  {author} {\bibinfo {author} {\bibfnamefont {F.~D.~M.}\
			\bibnamefont {Haldane}},\ }\href
	{https://doi.org/10.1103/PhysRevLett.93.206602} {\bibfield  {journal}
		{\bibinfo  {journal} {Phys. Rev. Lett.}\ }\textbf {\bibinfo {volume} {93}},\
		\bibinfo {pages} {206602} (\bibinfo {year} {2004})}\BibitemShut {NoStop}%
	\bibitem [{\citenamefont {Wang}\ \emph {et~al.}(2007)\citenamefont {Wang},
		\citenamefont {Vanderbilt}, \citenamefont {Yates},\ and\ \citenamefont
		{Souza}}]{PhysRevB.76.195109}%
	\BibitemOpen
	\bibfield  {author} {\bibinfo {author} {\bibfnamefont {X.}~\bibnamefont
			{Wang}}, \bibinfo {author} {\bibfnamefont {D.}~\bibnamefont {Vanderbilt}},
		\bibinfo {author} {\bibfnamefont {J.~R.}\ \bibnamefont {Yates}},\ and\
		\bibinfo {author} {\bibfnamefont {I.}~\bibnamefont {Souza}},\ }\href
	{https://doi.org/10.1103/PhysRevB.76.195109} {\bibfield  {journal} {\bibinfo
			{journal} {Phys. Rev. B}\ }\textbf {\bibinfo {volume} {76}},\ \bibinfo
		{pages} {195109} (\bibinfo {year} {2007})}\BibitemShut {NoStop}%
	\bibitem [{\citenamefont {Wang}\ \emph {et~al.}(2019)\citenamefont {Wang},
		\citenamefont {Wang}, \citenamefont {Min},\ and\ \citenamefont
		{Xia}}]{PhysRevB.99.224416}%
	\BibitemOpen
	\bibfield  {author} {\bibinfo {author} {\bibfnamefont {L.}~\bibnamefont
			{Wang}}, \bibinfo {author} {\bibfnamefont {X.~R.}\ \bibnamefont {Wang}},
		\bibinfo {author} {\bibfnamefont {T.}~\bibnamefont {Min}},\ and\ \bibinfo
		{author} {\bibfnamefont {K.}~\bibnamefont {Xia}},\ }\href
	{https://doi.org/10.1103/PhysRevB.99.224416} {\bibfield  {journal} {\bibinfo
			{journal} {Phys. Rev. B}\ }\textbf {\bibinfo {volume} {99}},\ \bibinfo
		{pages} {224416} (\bibinfo {year} {2019})}\BibitemShut {NoStop}%
	\bibitem [{\citenamefont {Besikci}\ \emph {et~al.}(1993)\citenamefont
		{Besikci}, \citenamefont {Choi}, \citenamefont {Sudharsanan},\ and\
		\citenamefont {Razeghi}}]{jap1.353821}%
	\BibitemOpen
	\bibfield  {author} {\bibinfo {author} {\bibfnamefont {C.}~\bibnamefont
			{Besikci}}, \bibinfo {author} {\bibfnamefont {Y.~H.}\ \bibnamefont {Choi}},
		\bibinfo {author} {\bibfnamefont {R.}~\bibnamefont {Sudharsanan}},\ and\
		\bibinfo {author} {\bibfnamefont {M.}~\bibnamefont {Razeghi}},\ }\href
	{https://doi.org/10.1063/1.353821} {\bibfield  {journal} {\bibinfo  {journal}
			{Journal of Applied Physics}\ }\textbf {\bibinfo {volume} {73}},\ \bibinfo
		{pages} {5009} (\bibinfo {year} {1993})}\BibitemShut {NoStop}%
	\bibitem [{\citenamefont {Gorini}\ \emph {et~al.}(2015)\citenamefont {Gorini},
		\citenamefont {Eckern},\ and\ \citenamefont
		{Raimondi}}]{PhysRevLett.115.076602}%
	\BibitemOpen
	\bibfield  {author} {\bibinfo {author} {\bibfnamefont {C.}~\bibnamefont
			{Gorini}}, \bibinfo {author} {\bibfnamefont {U.}~\bibnamefont {Eckern}},\
		and\ \bibinfo {author} {\bibfnamefont {R.}~\bibnamefont {Raimondi}},\ }\href
	{https://doi.org/10.1103/PhysRevLett.115.076602} {\bibfield  {journal}
		{\bibinfo  {journal} {Phys. Rev. Lett.}\ }\textbf {\bibinfo {volume} {115}},\
		\bibinfo {pages} {076602} (\bibinfo {year} {2015})}\BibitemShut {NoStop}%
	\bibitem [{\citenamefont {{Shitade}}\ and\ \citenamefont
		{{Nagaosa}}(2012)}]{2012JPSJ}%
	\BibitemOpen
	\bibfield  {author} {\bibinfo {author} {\bibfnamefont {A.}~\bibnamefont
			{{Shitade}}}\ and\ \bibinfo {author} {\bibfnamefont {N.}~\bibnamefont
			{{Nagaosa}}},\ }\href {https://doi.org/10.1143/JPSJ.81.083704} {\bibfield
		{journal} {\bibinfo  {journal} {Journal of the Physical Society of Japan}\
		}\textbf {\bibinfo {volume} {81}},\ \bibinfo {pages} {083704} (\bibinfo
		{year} {2012})},\ \Eprint {https://arxiv.org/abs/1109.5463} {arXiv:1109.5463
		[cond-mat.str-el]} \BibitemShut {NoStop}%
	\bibitem [{\citenamefont {Liu}\ \emph {et~al.}(2007)\citenamefont {Liu},
		\citenamefont {Horing},\ and\ \citenamefont {Lei}}]{PhysRevB.76.195309}%
	\BibitemOpen
	\bibfield  {author} {\bibinfo {author} {\bibfnamefont {S.~Y.}\ \bibnamefont
			{Liu}}, \bibinfo {author} {\bibfnamefont {N.~J.~M.}\ \bibnamefont {Horing}},\
		and\ \bibinfo {author} {\bibfnamefont {X.~L.}\ \bibnamefont {Lei}},\ }\href
	{https://doi.org/10.1103/PhysRevB.76.195309} {\bibfield  {journal} {\bibinfo
			{journal} {Phys. Rev. B}\ }\textbf {\bibinfo {volume} {76}},\ \bibinfo
		{pages} {195309} (\bibinfo {year} {2007})}\BibitemShut {NoStop}%
	\bibitem [{\citenamefont {Weischenberg}\ \emph {et~al.}(2011)\citenamefont
		{Weischenberg}, \citenamefont {Freimuth}, \citenamefont {Sinova},
		\citenamefont {Bl\"ugel},\ and\ \citenamefont
		{Mokrousov}}]{PhysRevLett.107.106601}%
	\BibitemOpen
	\bibfield  {author} {\bibinfo {author} {\bibfnamefont {J.}~\bibnamefont
			{Weischenberg}}, \bibinfo {author} {\bibfnamefont {F.}~\bibnamefont
			{Freimuth}}, \bibinfo {author} {\bibfnamefont {J.}~\bibnamefont {Sinova}},
		\bibinfo {author} {\bibfnamefont {S.}~\bibnamefont {Bl\"ugel}},\ and\
		\bibinfo {author} {\bibfnamefont {Y.}~\bibnamefont {Mokrousov}},\ }\href
	{https://doi.org/10.1103/PhysRevLett.107.106601} {\bibfield  {journal}
		{\bibinfo  {journal} {Phys. Rev. Lett.}\ }\textbf {\bibinfo {volume} {107}},\
		\bibinfo {pages} {106601} (\bibinfo {year} {2011})}\BibitemShut {NoStop}%
	\bibitem [{\citenamefont {Lowitzer}\ \emph {et~al.}(2010)\citenamefont
		{Lowitzer}, \citenamefont {K\"odderitzsch},\ and\ \citenamefont
		{Ebert}}]{PhysRevLett.105.266604}%
	\BibitemOpen
	\bibfield  {author} {\bibinfo {author} {\bibfnamefont {S.}~\bibnamefont
			{Lowitzer}}, \bibinfo {author} {\bibfnamefont {D.}~\bibnamefont
			{K\"odderitzsch}},\ and\ \bibinfo {author} {\bibfnamefont {H.}~\bibnamefont
			{Ebert}},\ }\href {https://doi.org/10.1103/PhysRevLett.105.266604} {\bibfield
		{journal} {\bibinfo  {journal} {Phys. Rev. Lett.}\ }\textbf {\bibinfo
			{volume} {105}},\ \bibinfo {pages} {266604} (\bibinfo {year}
		{2010})}\BibitemShut {NoStop}%
	\bibitem [{\citenamefont {Su}\ \emph {et~al.}(2014)\citenamefont {Su},
		\citenamefont {Li}, \citenamefont {Hou}, \citenamefont {Jin}, \citenamefont
		{Liu},\ and\ \citenamefont {Wang}}]{PhysRevB.90.214410}%
	\BibitemOpen
	\bibfield  {author} {\bibinfo {author} {\bibfnamefont {G.}~\bibnamefont
			{Su}}, \bibinfo {author} {\bibfnamefont {Y.}~\bibnamefont {Li}}, \bibinfo
		{author} {\bibfnamefont {D.}~\bibnamefont {Hou}}, \bibinfo {author}
		{\bibfnamefont {X.}~\bibnamefont {Jin}}, \bibinfo {author} {\bibfnamefont
			{H.}~\bibnamefont {Liu}},\ and\ \bibinfo {author} {\bibfnamefont
			{S.}~\bibnamefont {Wang}},\ }\href
	{https://doi.org/10.1103/PhysRevB.90.214410} {\bibfield  {journal} {\bibinfo
			{journal} {Phys. Rev. B}\ }\textbf {\bibinfo {volume} {90}},\ \bibinfo
		{pages} {214410} (\bibinfo {year} {2014})}\BibitemShut {NoStop}%
	\bibitem [{\citenamefont {Grigoryan}\ \emph {et~al.}(2017)\citenamefont
		{Grigoryan}, \citenamefont {Xiao}, \citenamefont {Wang},\ and\ \citenamefont
		{Xia}}]{Grigoryan2016Scaling}%
	\BibitemOpen
	\bibfield  {author} {\bibinfo {author} {\bibfnamefont {V.~L.}\ \bibnamefont
			{Grigoryan}}, \bibinfo {author} {\bibfnamefont {J.}~\bibnamefont {Xiao}},
		\bibinfo {author} {\bibfnamefont {X.}~\bibnamefont {Wang}},\ and\ \bibinfo
		{author} {\bibfnamefont {K.}~\bibnamefont {Xia}},\ }\href
	{https://doi.org/10.1103/PhysRevB.96.144426} {\bibfield  {journal} {\bibinfo
			{journal} {Phys. Rev. B}\ }\textbf {\bibinfo {volume} {96}},\ \bibinfo
		{pages} {144426} (\bibinfo {year} {2017})}\BibitemShut {NoStop}%
	\bibitem [{\citenamefont {Yang}\ \emph {et~al.}(2011)\citenamefont {Yang},
		\citenamefont {Pan}, \citenamefont {Yao},\ and\ \citenamefont
		{Niu}}]{PhysRevB.83.125122}%
	\BibitemOpen
	\bibfield  {author} {\bibinfo {author} {\bibfnamefont {S.~A.}\ \bibnamefont
			{Yang}}, \bibinfo {author} {\bibfnamefont {H.}~\bibnamefont {Pan}}, \bibinfo
		{author} {\bibfnamefont {Y.}~\bibnamefont {Yao}},\ and\ \bibinfo {author}
		{\bibfnamefont {Q.}~\bibnamefont {Niu}},\ }\href
	{https://doi.org/10.1103/PhysRevB.83.125122} {\bibfield  {journal} {\bibinfo
			{journal} {Phys. Rev. B}\ }\textbf {\bibinfo {volume} {83}},\ \bibinfo
		{pages} {125122} (\bibinfo {year} {2011})}\BibitemShut {NoStop}%
	\bibitem [{\citenamefont {Hou}\ \emph {et~al.}(2015)\citenamefont {Hou},
		\citenamefont {Su}, \citenamefont {Tian}, \citenamefont {Jin}, \citenamefont
		{Yang},\ and\ \citenamefont {Niu}}]{PhysRevLett.114.217203}%
	\BibitemOpen
	\bibfield  {author} {\bibinfo {author} {\bibfnamefont {D.}~\bibnamefont
			{Hou}}, \bibinfo {author} {\bibfnamefont {G.}~\bibnamefont {Su}}, \bibinfo
		{author} {\bibfnamefont {Y.}~\bibnamefont {Tian}}, \bibinfo {author}
		{\bibfnamefont {X.}~\bibnamefont {Jin}}, \bibinfo {author} {\bibfnamefont
			{S.~A.}\ \bibnamefont {Yang}},\ and\ \bibinfo {author} {\bibfnamefont
			{Q.}~\bibnamefont {Niu}},\ }\href
	{https://doi.org/10.1103/PhysRevLett.114.217203} {\bibfield  {journal}
		{\bibinfo  {journal} {Phys. Rev. Lett.}\ }\textbf {\bibinfo {volume} {114}},\
		\bibinfo {pages} {217203} (\bibinfo {year} {2015})}\BibitemShut {NoStop}%
	\bibitem [{\citenamefont {Ye}\ \emph {et~al.}(2012)\citenamefont {Ye},
		\citenamefont {Tian}, \citenamefont {Jin},\ and\ \citenamefont
		{Xiao}}]{PhysRevB.85.220403}%
	\BibitemOpen
	\bibfield  {author} {\bibinfo {author} {\bibfnamefont {L.}~\bibnamefont
			{Ye}}, \bibinfo {author} {\bibfnamefont {Y.}~\bibnamefont {Tian}}, \bibinfo
		{author} {\bibfnamefont {X.}~\bibnamefont {Jin}},\ and\ \bibinfo {author}
		{\bibfnamefont {D.}~\bibnamefont {Xiao}},\ }\href
	{https://doi.org/10.1103/PhysRevB.85.220403} {\bibfield  {journal} {\bibinfo
			{journal} {Phys. Rev. B}\ }\textbf {\bibinfo {volume} {85}},\ \bibinfo
		{pages} {220403} (\bibinfo {year} {2012})}\BibitemShut {NoStop}%
	\bibitem [{\citenamefont {Xia}\ \emph {et~al.}(2006)\citenamefont {Xia},
		\citenamefont {Zwierzycki}, \citenamefont {Talanana}, \citenamefont {Kelly},\
		and\ \citenamefont {Bauer}}]{PhysRevB.73.064420}%
	\BibitemOpen
	\bibfield  {author} {\bibinfo {author} {\bibfnamefont {K.}~\bibnamefont
			{Xia}}, \bibinfo {author} {\bibfnamefont {M.}~\bibnamefont {Zwierzycki}},
		\bibinfo {author} {\bibfnamefont {M.}~\bibnamefont {Talanana}}, \bibinfo
		{author} {\bibfnamefont {P.~J.}\ \bibnamefont {Kelly}},\ and\ \bibinfo
		{author} {\bibfnamefont {G.~E.~W.}\ \bibnamefont {Bauer}},\ }\href
	{https://doi.org/10.1103/PhysRevB.73.064420} {\bibfield  {journal} {\bibinfo
			{journal} {Phys. Rev. B}\ }\textbf {\bibinfo {volume} {73}},\ \bibinfo
		{pages} {064420} (\bibinfo {year} {2006})}\BibitemShut {NoStop}%
	\bibitem [{\citenamefont {Wang}\ \emph {et~al.}(2008)\citenamefont {Wang},
		\citenamefont {Xu},\ and\ \citenamefont {Xia}}]{PhysRevB.77.184430}%
	\BibitemOpen
	\bibfield  {author} {\bibinfo {author} {\bibfnamefont {S.}~\bibnamefont
			{Wang}}, \bibinfo {author} {\bibfnamefont {Y.}~\bibnamefont {Xu}},\ and\
		\bibinfo {author} {\bibfnamefont {K.}~\bibnamefont {Xia}},\ }\href
	{https://doi.org/10.1103/PhysRevB.77.184430} {\bibfield  {journal} {\bibinfo
			{journal} {Phys. Rev. B}\ }\textbf {\bibinfo {volume} {77}},\ \bibinfo
		{pages} {184430} (\bibinfo {year} {2008})}\BibitemShut {NoStop}%
	\bibitem [{\citenamefont {Wang}\ \emph {et~al.}(2016)\citenamefont {Wang},
		\citenamefont {Wesselink}, \citenamefont {Liu}, \citenamefont {Yuan},
		\citenamefont {Xia},\ and\ \citenamefont {Kelly}}]{Wang.2016}%
	\BibitemOpen
	\bibfield  {author} {\bibinfo {author} {\bibfnamefont {L.}~\bibnamefont
			{Wang}}, \bibinfo {author} {\bibfnamefont {R.~J.~H.}\ \bibnamefont
			{Wesselink}}, \bibinfo {author} {\bibfnamefont {Y.}~\bibnamefont {Liu}},
		\bibinfo {author} {\bibfnamefont {Z.}~\bibnamefont {Yuan}}, \bibinfo {author}
		{\bibfnamefont {K.}~\bibnamefont {Xia}},\ and\ \bibinfo {author}
		{\bibfnamefont {P.~J.}\ \bibnamefont {Kelly}},\ }\href
	{https://doi.org/10.1103/PhysRevLett.116.196602} {\bibfield  {journal}
		{\bibinfo  {journal} {Physical review letters}\ }\textbf {\bibinfo {volume}
			{116}},\ \bibinfo {pages} {196602} (\bibinfo {year} {2016})}\BibitemShut
	{NoStop}%
	\bibitem [{\citenamefont {Starikov}\ \emph {et~al.}(2018)\citenamefont
		{Starikov}, \citenamefont {Liu}, \citenamefont {Yuan},\ and\ \citenamefont
		{Kelly}}]{PhysRevB.97.214415}%
	\BibitemOpen
	\bibfield  {author} {\bibinfo {author} {\bibfnamefont {A.~A.}\ \bibnamefont
			{Starikov}}, \bibinfo {author} {\bibfnamefont {Y.}~\bibnamefont {Liu}},
		\bibinfo {author} {\bibfnamefont {Z.}~\bibnamefont {Yuan}},\ and\ \bibinfo
		{author} {\bibfnamefont {P.~J.}\ \bibnamefont {Kelly}},\ }\href
	{https://doi.org/10.1103/PhysRevB.97.214415} {\bibfield  {journal} {\bibinfo
			{journal} {Phys. Rev. B}\ }\textbf {\bibinfo {volume} {97}},\ \bibinfo
		{pages} {214415} (\bibinfo {year} {2018})}\BibitemShut {NoStop}%
	\bibitem [{\citenamefont {Wesselink}\ \emph {et~al.}(2019)\citenamefont
		{Wesselink}, \citenamefont {Gupta}, \citenamefont {Yuan},\ and\ \citenamefont
		{Kelly}}]{PhysRevB.99.144409}%
	\BibitemOpen
	\bibfield  {author} {\bibinfo {author} {\bibfnamefont {R.~J.~H.}\
			\bibnamefont {Wesselink}}, \bibinfo {author} {\bibfnamefont {K.}~\bibnamefont
			{Gupta}}, \bibinfo {author} {\bibfnamefont {Z.}~\bibnamefont {Yuan}},\ and\
		\bibinfo {author} {\bibfnamefont {P.~J.}\ \bibnamefont {Kelly}},\ }\href
	{https://doi.org/10.1103/PhysRevB.99.144409} {\bibfield  {journal} {\bibinfo
			{journal} {Phys. Rev. B}\ }\textbf {\bibinfo {volume} {99}},\ \bibinfo
		{pages} {144409} (\bibinfo {year} {2019})}\BibitemShut {NoStop}%
	\bibitem [{\citenamefont {Li}\ \emph {et~al.}(2019)\citenamefont {Li},
		\citenamefont {Shen},\ and\ \citenamefont {Xia}}]{PhysRevB.99.134427}%
	\BibitemOpen
	\bibfield  {author} {\bibinfo {author} {\bibfnamefont {S.}~\bibnamefont
			{Li}}, \bibinfo {author} {\bibfnamefont {K.}~\bibnamefont {Shen}},\ and\
		\bibinfo {author} {\bibfnamefont {K.}~\bibnamefont {Xia}},\ }\href
	{https://doi.org/10.1103/PhysRevB.99.134427} {\bibfield  {journal} {\bibinfo
			{journal} {Phys. Rev. B}\ }\textbf {\bibinfo {volume} {99}},\ \bibinfo
		{pages} {134427} (\bibinfo {year} {2019})}\BibitemShut {NoStop}%
	\bibitem [{\citenamefont {{Andersen}}\ \emph {et~al.}(1995)\citenamefont
		{{Andersen}}, \citenamefont {{Jepsen}},\ and\ \citenamefont
		{{Krier}}}]{andersen1995exact}%
	\BibitemOpen
	\bibfield  {author} {\bibinfo {author} {\bibfnamefont {O.~K.}\ \bibnamefont
			{{Andersen}}}, \bibinfo {author} {\bibfnamefont {O.}~\bibnamefont
			{{Jepsen}}},\ and\ \bibinfo {author} {\bibfnamefont {G.}~\bibnamefont
			{{Krier}}},\ }in\ \href {https://academic.microsoft.com/paper/2331546596}
	{\emph {\bibinfo {booktitle} {Proceedings of the Miniworkshop on Methods of
				Electronic Structure Calculations and Working Group on Disordered Alloys}}}\
	(\bibinfo {year} {1995})\ pp.\ \bibinfo {pages} {63--124}\BibitemShut
	{NoStop}%
	\bibitem [{\citenamefont {Vitos}(2001)}]{PhysRevB.64.014107}%
	\BibitemOpen
	\bibfield  {author} {\bibinfo {author} {\bibfnamefont {L.}~\bibnamefont
			{Vitos}},\ }\href {https://doi.org/10.1103/PhysRevB.64.014107} {\bibfield
		{journal} {\bibinfo  {journal} {Phys. Rev. B}\ }\textbf {\bibinfo {volume}
			{64}},\ \bibinfo {pages} {014107} (\bibinfo {year} {2001})}\BibitemShut
	{NoStop}%
	\bibitem [{\citenamefont {Vitos}\ \emph {et~al.}(2000)\citenamefont {Vitos},
		\citenamefont {Skriver}, \citenamefont {Johansson},\ and\ \citenamefont
		{Koll\'{a}r}}]{VITOS200024}%
	\BibitemOpen
	\bibfield  {author} {\bibinfo {author} {\bibfnamefont {L.}~\bibnamefont
			{Vitos}}, \bibinfo {author} {\bibfnamefont {H.}~\bibnamefont {Skriver}},
		\bibinfo {author} {\bibfnamefont {B.}~\bibnamefont {Johansson}},\ and\
		\bibinfo {author} {\bibfnamefont {J.}~\bibnamefont {Koll\'{a}r}},\ }\href
	{https://doi.org/https://doi.org/10.1016/S0927-0256(99)00098-1} {\bibfield
		{journal} {\bibinfo  {journal} {Computational Materials Science}\ }\textbf
		{\bibinfo {volume} {18}},\ \bibinfo {pages} {24 } (\bibinfo {year}
		{2000})}\BibitemShut {NoStop}%
	\bibitem [{\citenamefont {Pourovskii}\ \emph {et~al.}(2005)\citenamefont
		{Pourovskii}, \citenamefont {Ruban}, \citenamefont {Vitos}, \citenamefont
		{Ebert}, \citenamefont {Johansson},\ and\ \citenamefont
		{Abrikosov}}]{PhysRevB.71.094415}%
	\BibitemOpen
	\bibfield  {author} {\bibinfo {author} {\bibfnamefont {L.~V.}\ \bibnamefont
			{Pourovskii}}, \bibinfo {author} {\bibfnamefont {A.~V.}\ \bibnamefont
			{Ruban}}, \bibinfo {author} {\bibfnamefont {L.}~\bibnamefont {Vitos}},
		\bibinfo {author} {\bibfnamefont {H.}~\bibnamefont {Ebert}}, \bibinfo
		{author} {\bibfnamefont {B.}~\bibnamefont {Johansson}},\ and\ \bibinfo
		{author} {\bibfnamefont {I.~A.}\ \bibnamefont {Abrikosov}},\ }\href
	{https://doi.org/10.1103/PhysRevB.71.094415} {\bibfield  {journal} {\bibinfo
			{journal} {Phys. Rev. B}\ }\textbf {\bibinfo {volume} {71}},\ \bibinfo
		{pages} {094415} (\bibinfo {year} {2005})}\BibitemShut {NoStop}%
	\bibitem [{\citenamefont {Vitos}\ \emph {et~al.}(2001)\citenamefont {Vitos},
		\citenamefont {Abrikosov},\ and\ \citenamefont
		{Johansson}}]{PhysRevLett.87.156401}%
	\BibitemOpen
	\bibfield  {author} {\bibinfo {author} {\bibfnamefont {L.}~\bibnamefont
			{Vitos}}, \bibinfo {author} {\bibfnamefont {I.~A.}\ \bibnamefont
			{Abrikosov}},\ and\ \bibinfo {author} {\bibfnamefont {B.}~\bibnamefont
			{Johansson}},\ }\href {https://doi.org/10.1103/PhysRevLett.87.156401}
	{\bibfield  {journal} {\bibinfo  {journal} {Phys. Rev. Lett.}\ }\textbf
		{\bibinfo {volume} {87}},\ \bibinfo {pages} {156401} (\bibinfo {year}
		{2001})}\BibitemShut {NoStop}%
	\bibitem [{\citenamefont {Tian}\ \emph {et~al.}(2013)\citenamefont {Tian},
		\citenamefont {Varga}, \citenamefont {Chen}, \citenamefont {Delczeg},\ and\
		\citenamefont {Vitos}}]{PhysRevB.87.075144}%
	\BibitemOpen
	\bibfield  {author} {\bibinfo {author} {\bibfnamefont {F.}~\bibnamefont
			{Tian}}, \bibinfo {author} {\bibfnamefont {L.~K.}\ \bibnamefont {Varga}},
		\bibinfo {author} {\bibfnamefont {N.}~\bibnamefont {Chen}}, \bibinfo {author}
		{\bibfnamefont {L.}~\bibnamefont {Delczeg}},\ and\ \bibinfo {author}
		{\bibfnamefont {L.}~\bibnamefont {Vitos}},\ }\href
	{https://doi.org/10.1103/PhysRevB.87.075144} {\bibfield  {journal} {\bibinfo
			{journal} {Phys. Rev. B}\ }\textbf {\bibinfo {volume} {87}},\ \bibinfo
		{pages} {075144} (\bibinfo {year} {2013})}\BibitemShut {NoStop}%
	\bibitem [{\citenamefont {Vitos}\ \emph {et~al.}(1998)\citenamefont {Vitos},
		\citenamefont {Ruban}, \citenamefont {Skriver},\ and\ \citenamefont
		{Koll\'{a}r}}]{VITOS1998186}%
	\BibitemOpen
	\bibfield  {author} {\bibinfo {author} {\bibfnamefont {L.}~\bibnamefont
			{Vitos}}, \bibinfo {author} {\bibfnamefont {A.}~\bibnamefont {Ruban}},
		\bibinfo {author} {\bibfnamefont {H.}~\bibnamefont {Skriver}},\ and\ \bibinfo
		{author} {\bibfnamefont {J.}~\bibnamefont {Koll\'{a}r}},\ }\href
	{https://doi.org/https://doi.org/10.1016/S0039-6028(98)00363-X} {\bibfield
		{journal} {\bibinfo  {journal} {Surface Science}\ }\textbf {\bibinfo {volume}
			{411}},\ \bibinfo {pages} {186 } (\bibinfo {year} {1998})}\BibitemShut
	{NoStop}%
	\bibitem [{\citenamefont {Chen}\ \emph {et~al.}(2020)\citenamefont {Chen},
		\citenamefont {Zhang}, \citenamefont {Zhang}, \citenamefont {Wang},
		\citenamefont {Sang},\ and\ \citenamefont {Ke}}]{PhysRevB.102.035405}%
	\BibitemOpen
	\bibfield  {author} {\bibinfo {author} {\bibfnamefont {Z.}~\bibnamefont
			{Chen}}, \bibinfo {author} {\bibfnamefont {Q.}~\bibnamefont {Zhang}},
		\bibinfo {author} {\bibfnamefont {Y.}~\bibnamefont {Zhang}}, \bibinfo
		{author} {\bibfnamefont {L.}~\bibnamefont {Wang}}, \bibinfo {author}
		{\bibfnamefont {M.}~\bibnamefont {Sang}},\ and\ \bibinfo {author}
		{\bibfnamefont {Y.}~\bibnamefont {Ke}},\ }\href
	{https://doi.org/10.1103/PhysRevB.102.035405} {\bibfield  {journal} {\bibinfo
			{journal} {Phys. Rev. B}\ }\textbf {\bibinfo {volume} {102}},\ \bibinfo
		{pages} {035405} (\bibinfo {year} {2020})}\BibitemShut {NoStop}%
	\bibitem [{\citenamefont {Zhang}\ \emph {et~al.}(2019)\citenamefont {Zhang},
		\citenamefont {Yan}, \citenamefont {Zhang},\ and\ \citenamefont
		{Ke}}]{PhysRevB.100.075134}%
	\BibitemOpen
	\bibfield  {author} {\bibinfo {author} {\bibfnamefont {Q.}~\bibnamefont
			{Zhang}}, \bibinfo {author} {\bibfnamefont {J.}~\bibnamefont {Yan}}, \bibinfo
		{author} {\bibfnamefont {Y.}~\bibnamefont {Zhang}},\ and\ \bibinfo {author}
		{\bibfnamefont {Y.}~\bibnamefont {Ke}},\ }\href
	{https://doi.org/10.1103/PhysRevB.100.075134} {\bibfield  {journal} {\bibinfo
			{journal} {Phys. Rev. B}\ }\textbf {\bibinfo {volume} {100}},\ \bibinfo
		{pages} {075134} (\bibinfo {year} {2019})}\BibitemShut {NoStop}%
	\bibitem [{\citenamefont {B\"uttiker}\ \emph {et~al.}(1985)\citenamefont
		{B\"uttiker}, \citenamefont {Imry}, \citenamefont {Landauer},\ and\
		\citenamefont {Pinhas}}]{PhysRevB.31.6207}%
	\BibitemOpen
	\bibfield  {author} {\bibinfo {author} {\bibfnamefont {M.}~\bibnamefont
			{B\"uttiker}}, \bibinfo {author} {\bibfnamefont {Y.}~\bibnamefont {Imry}},
		\bibinfo {author} {\bibfnamefont {R.}~\bibnamefont {Landauer}},\ and\
		\bibinfo {author} {\bibfnamefont {S.}~\bibnamefont {Pinhas}},\ }\href
	{https://doi.org/10.1103/PhysRevB.31.6207} {\bibfield  {journal} {\bibinfo
			{journal} {Phys. Rev. B}\ }\textbf {\bibinfo {volume} {31}},\ \bibinfo
		{pages} {6207} (\bibinfo {year} {1985})}\BibitemShut {NoStop}%
	\bibitem [{\citenamefont {Datta}(1992)}]{PhysRevB.45.1347}%
	\BibitemOpen
	\bibfield  {author} {\bibinfo {author} {\bibfnamefont {S.}~\bibnamefont
			{Datta}},\ }\href {https://doi.org/10.1103/PhysRevB.45.1347} {\bibfield
		{journal} {\bibinfo  {journal} {Phys. Rev. B}\ }\textbf {\bibinfo {volume}
			{45}},\ \bibinfo {pages} {1347} (\bibinfo {year} {1992})}\BibitemShut
	{NoStop}%
	\bibitem [{\citenamefont {Baranger}\ \emph {et~al.}(1991)\citenamefont
		{Baranger}, \citenamefont {DiVincenzo}, \citenamefont {Jalabert},\ and\
		\citenamefont {Stone}}]{PhysRevB.44.10637}%
	\BibitemOpen
	\bibfield  {author} {\bibinfo {author} {\bibfnamefont {H.~U.}\ \bibnamefont
			{Baranger}}, \bibinfo {author} {\bibfnamefont {D.~P.}\ \bibnamefont
			{DiVincenzo}}, \bibinfo {author} {\bibfnamefont {R.~A.}\ \bibnamefont
			{Jalabert}},\ and\ \bibinfo {author} {\bibfnamefont {A.~D.}\ \bibnamefont
			{Stone}},\ }\href {https://doi.org/10.1103/PhysRevB.44.10637} {\bibfield
		{journal} {\bibinfo  {journal} {Phys. Rev. B}\ }\textbf {\bibinfo {volume}
			{44}},\ \bibinfo {pages} {10637} (\bibinfo {year} {1991})}\BibitemShut
	{NoStop}%
	\bibitem [{\citenamefont {Datta}(1995)}]{datta_1995}%
	\BibitemOpen
	\bibfield  {author} {\bibinfo {author} {\bibfnamefont {S.}~\bibnamefont
			{Datta}},\ }\href {https://doi.org/10.1017/CBO9780511805776} {\emph {\bibinfo
			{title} {Electronic Transport in Mesoscopic Systems}}},\ Cambridge Studies in
	Semiconductor Physics and Microelectronic Engineering\ (\bibinfo  {publisher}
	{Cambridge University Press},\ \bibinfo {year} {1995})\BibitemShut {NoStop}%
	\bibitem [{\citenamefont {Imry}(2002)}]{Imry_2002}%
	\BibitemOpen
	\bibfield  {author} {\bibinfo {author} {\bibfnamefont {Y.}~\bibnamefont
			{Imry}},\ }\href
	{https://global.oup.com/academic/product/introduction-to-mesoscopic-physics-9780195101676?cc=cn&lang=en&}
	{\emph {\bibinfo {title} {{Introduction to mesoscopic physics}}}},\
	Mesoscopic physics and nanotechnology\ (\bibinfo  {publisher} {Oxford
		University Press},\ \bibinfo {address} {New York, NY},\ \bibinfo {year}
	{2002})\BibitemShut {NoStop}%
	\bibitem [{\citenamefont {{Vitos}}(2007)}]{vitos2007computational}%
	\BibitemOpen
	\bibfield  {author} {\bibinfo {author} {\bibfnamefont {L.}~\bibnamefont
			{{Vitos}}},\ }\href {https://academic.microsoft.com/paper/644091946} {\emph
		{\bibinfo {title} {Computational Quantum Mechanics for Materials Engineers:
				The EMTO Method and Applications}}}\ (\bibinfo {year} {2007})\BibitemShut
	{NoStop}%
	\bibitem [{emt()}]{emto}%
	\BibitemOpen
	\href@noop {} {\ }\bibinfo {note} {The open source EMTO-CPA code can be
		obtained at http://emto.gitlab.io/index.html}\BibitemShut {NoStop}%
	\bibitem [{\citenamefont {Ando}(1991)}]{PhysRevB.44.8017}%
	\BibitemOpen
	\bibfield  {author} {\bibinfo {author} {\bibfnamefont {T.}~\bibnamefont
			{Ando}},\ }\href {https://doi.org/10.1103/PhysRevB.44.8017} {\bibfield
		{journal} {\bibinfo  {journal} {Phys. Rev. B}\ }\textbf {\bibinfo {volume}
			{44}},\ \bibinfo {pages} {8017} (\bibinfo {year} {1991})}\BibitemShut
	{NoStop}%
	\bibitem [{\citenamefont {Turek}\ \emph {et~al.}(2002)\citenamefont {Turek},
		\citenamefont {Kudrnovsk\'y}, \citenamefont {Drchal}, \citenamefont
		{Szunyogh},\ and\ \citenamefont {Weinberger}}]{PhysRevB.65.125101}%
	\BibitemOpen
	\bibfield  {author} {\bibinfo {author} {\bibfnamefont {I.}~\bibnamefont
			{Turek}}, \bibinfo {author} {\bibfnamefont {J.}~\bibnamefont {Kudrnovsk\'y}},
		\bibinfo {author} {\bibfnamefont {V.}~\bibnamefont {Drchal}}, \bibinfo
		{author} {\bibfnamefont {L.}~\bibnamefont {Szunyogh}},\ and\ \bibinfo
		{author} {\bibfnamefont {P.}~\bibnamefont {Weinberger}},\ }\href
	{https://doi.org/10.1103/PhysRevB.65.125101} {\bibfield  {journal} {\bibinfo
			{journal} {Phys. Rev. B}\ }\textbf {\bibinfo {volume} {65}},\ \bibinfo
		{pages} {125101} (\bibinfo {year} {2002})}\BibitemShut {NoStop}%
	\bibitem [{\citenamefont {Ziman}(1960)}]{ziman1960}%
	\BibitemOpen
	\bibfield  {author} {\bibinfo {author} {\bibfnamefont {J.}~\bibnamefont
			{Ziman}},\ }\href {https://doi.org/10.1063/1.3057244} {\emph {\bibinfo
			{title} {Electrons and Phonons}}},\ Vol.~\bibinfo {volume} {14}\ (\bibinfo
	{publisher} {Oxford},\ \bibinfo {year} {1960})\BibitemShut {NoStop}%
	\bibitem [{\citenamefont {Zhao}\ \emph {et~al.}(2011)\citenamefont {Zhao},
		\citenamefont {Qu},\ and\ \citenamefont {Xia}}]{jap1.3638694}%
	\BibitemOpen
	\bibfield  {author} {\bibinfo {author} {\bibfnamefont {Y.-N.}\ \bibnamefont
			{Zhao}}, \bibinfo {author} {\bibfnamefont {S.-X.}\ \bibnamefont {Qu}},\ and\
		\bibinfo {author} {\bibfnamefont {K.}~\bibnamefont {Xia}},\ }\href
	{http://scitation.aip.org/content/aip/journal/jap/110/6/10.1063/1.3638694}
	{\bibfield  {journal} {\bibinfo  {journal} {Journal of Applied Physics}\
		}\textbf {\bibinfo {volume} {110}},\ \bibinfo {eid} {064312} (\bibinfo {year}
		{2011})}\BibitemShut {NoStop}%
	\bibitem [{\citenamefont {Liu}\ \emph {et~al.}(2011)\citenamefont {Liu},
		\citenamefont {Starikov}, \citenamefont {Yuan},\ and\ \citenamefont
		{Kelly}}]{PhysRevB.84.014412}%
	\BibitemOpen
	\bibfield  {author} {\bibinfo {author} {\bibfnamefont {Y.}~\bibnamefont
			{Liu}}, \bibinfo {author} {\bibfnamefont {A.~A.}\ \bibnamefont {Starikov}},
		\bibinfo {author} {\bibfnamefont {Z.}~\bibnamefont {Yuan}},\ and\ \bibinfo
		{author} {\bibfnamefont {P.~J.}\ \bibnamefont {Kelly}},\ }\href
	{https://doi.org/10.1103/PhysRevB.84.014412} {\bibfield  {journal} {\bibinfo
			{journal} {Phys. Rev. B}\ }\textbf {\bibinfo {volume} {84}},\ \bibinfo
		{pages} {014412} (\bibinfo {year} {2011})}\BibitemShut {NoStop}%
	\bibitem [{\citenamefont {Liu}\ \emph {et~al.}(2015)\citenamefont {Liu},
		\citenamefont {Yuan}, \citenamefont {Wesselink}, \citenamefont {Starikov},
		\citenamefont {van Schilfgaarde},\ and\ \citenamefont
		{Kelly}}]{PhysRevB.91.220405}%
	\BibitemOpen
	\bibfield  {author} {\bibinfo {author} {\bibfnamefont {Y.}~\bibnamefont
			{Liu}}, \bibinfo {author} {\bibfnamefont {Z.}~\bibnamefont {Yuan}}, \bibinfo
		{author} {\bibfnamefont {R.~J.~H.}\ \bibnamefont {Wesselink}}, \bibinfo
		{author} {\bibfnamefont {A.~A.}\ \bibnamefont {Starikov}}, \bibinfo {author}
		{\bibfnamefont {M.}~\bibnamefont {van Schilfgaarde}},\ and\ \bibinfo {author}
		{\bibfnamefont {P.~J.}\ \bibnamefont {Kelly}},\ }\href
	{https://doi.org/10.1103/PhysRevB.91.220405} {\bibfield  {journal} {\bibinfo
			{journal} {Phys. Rev. B}\ }\textbf {\bibinfo {volume} {91}},\ \bibinfo
		{pages} {220405} (\bibinfo {year} {2015})}\BibitemShut {NoStop}%
	\bibitem [{\citenamefont {Kittel}(2005)}]{Kittel2005}%
	\BibitemOpen
	\bibfield  {author} {\bibinfo {author} {\bibfnamefont {C.}~\bibnamefont
			{Kittel}},\ }\href
	{http://www.wiley.com/WileyCDA/WileyTitle/productCd-EHEP000803.html} {\emph
		{\bibinfo {title} {Introduction To Solid State Physics, 8Th Edition}}}\
	(\bibinfo {year} {2005})\BibitemShut {NoStop}%
	\bibitem [{\citenamefont {Tian}\ \emph {et~al.}(2009)\citenamefont {Tian},
		\citenamefont {Ye},\ and\ \citenamefont {Jin}}]{PhysRevLett.103.087206}%
	\BibitemOpen
	\bibfield  {author} {\bibinfo {author} {\bibfnamefont {Y.}~\bibnamefont
			{Tian}}, \bibinfo {author} {\bibfnamefont {L.}~\bibnamefont {Ye}},\ and\
		\bibinfo {author} {\bibfnamefont {X.}~\bibnamefont {Jin}},\ }\href
	{https://doi.org/10.1103/PhysRevLett.103.087206} {\bibfield  {journal}
		{\bibinfo  {journal} {Phys. Rev. Lett.}\ }\textbf {\bibinfo {volume} {103}},\
		\bibinfo {pages} {087206} (\bibinfo {year} {2009})}\BibitemShut {NoStop}%
	\bibitem [{\citenamefont {Hou}\ \emph {et~al.}(2012)\citenamefont {Hou},
		\citenamefont {Li}, \citenamefont {Wei}, \citenamefont {Tian}, \citenamefont
		{Wu},\ and\ \citenamefont {Jin}}]{Hou_2012}%
	\BibitemOpen
	\bibfield  {author} {\bibinfo {author} {\bibfnamefont {D.}~\bibnamefont
			{Hou}}, \bibinfo {author} {\bibfnamefont {Y.}~\bibnamefont {Li}}, \bibinfo
		{author} {\bibfnamefont {D.}~\bibnamefont {Wei}}, \bibinfo {author}
		{\bibfnamefont {D.}~\bibnamefont {Tian}}, \bibinfo {author} {\bibfnamefont
			{L.}~\bibnamefont {Wu}},\ and\ \bibinfo {author} {\bibfnamefont
			{X.}~\bibnamefont {Jin}},\ }\href
	{https://doi.org/10.1088/0953-8984/24/48/482001} {\bibfield  {journal}
		{\bibinfo  {journal} {Journal of Physics: Condensed Matter}\ }\textbf
		{\bibinfo {volume} {24}},\ \bibinfo {pages} {482001} (\bibinfo {year}
		{2012})}\BibitemShut {NoStop}%
	\bibitem [{\citenamefont {Zhu}\ \emph {et~al.}(2014)\citenamefont {Zhu},
		\citenamefont {Pan},\ and\ \citenamefont {Zhao}}]{PhysRevB.89.220406}%
	\BibitemOpen
	\bibfield  {author} {\bibinfo {author} {\bibfnamefont {L.~J.}\ \bibnamefont
			{Zhu}}, \bibinfo {author} {\bibfnamefont {D.}~\bibnamefont {Pan}},\ and\
		\bibinfo {author} {\bibfnamefont {J.~H.}\ \bibnamefont {Zhao}},\ }\href
	{https://doi.org/10.1103/PhysRevB.89.220406} {\bibfield  {journal} {\bibinfo
			{journal} {Phys. Rev. B}\ }\textbf {\bibinfo {volume} {89}},\ \bibinfo
		{pages} {220406} (\bibinfo {year} {2014})}\BibitemShut {NoStop}%
	\bibitem [{\citenamefont {Shitade}\ and\ \citenamefont
		{Nagaosa}(2012)}]{doi:10.1143/JPSJ.81.083704}%
	\BibitemOpen
	\bibfield  {author} {\bibinfo {author} {\bibfnamefont {A.}~\bibnamefont
			{Shitade}}\ and\ \bibinfo {author} {\bibfnamefont {N.}~\bibnamefont
			{Nagaosa}},\ }\href {https://doi.org/10.1143/JPSJ.81.083704} {\bibfield
		{journal} {\bibinfo  {journal} {Journal of the Physical Society of Japan}\
		}\textbf {\bibinfo {volume} {81}},\ \bibinfo {pages} {083704} (\bibinfo
		{year} {2012})}\BibitemShut {NoStop}%
	\bibitem [{\citenamefont {Shiomi}(2013)}]{Shiomi2013}%
	\BibitemOpen
	\bibfield  {author} {\bibinfo {author} {\bibfnamefont {Y.}~\bibnamefont
			{Shiomi}},\ }\bibinfo {title} {Skew-scattering-induced anomalous hall effect
		in impurity-doped fe},\ in\ \href
	{https://doi.org/10.1007/978-4-431-54361-9_4} {\emph {\bibinfo {booktitle}
			{Anomalous and Topological Hall Effects in Itinerant Magnets}}}\ (\bibinfo
	{publisher} {Springer Japan},\ \bibinfo {address} {Tokyo},\ \bibinfo {year}
	{2013})\ pp.\ \bibinfo {pages} {47--63}\BibitemShut {NoStop}%
	\bibitem [{\citenamefont {St\ifmmode~\check{r}\else
			\v{r}\fi{}eda}(2013)}]{PhysRevB.88.134422}%
	\BibitemOpen
	\bibfield  {author} {\bibinfo {author} {\bibfnamefont {P.}~\bibnamefont
			{St\ifmmode~\check{r}\else \v{r}\fi{}eda}},\ }\href
	{https://doi.org/10.1103/PhysRevB.88.134422} {\bibfield  {journal} {\bibinfo
			{journal} {Phys. Rev. B}\ }\textbf {\bibinfo {volume} {88}},\ \bibinfo
		{pages} {134422} (\bibinfo {year} {2013})}\BibitemShut {NoStop}%
	\bibitem [{\citenamefont {Dheer}(1967)}]{PhysRev.156.637}%
	\BibitemOpen
	\bibfield  {author} {\bibinfo {author} {\bibfnamefont {P.~N.}\ \bibnamefont
			{Dheer}},\ }\href {https://doi.org/10.1103/PhysRev.156.637} {\bibfield
		{journal} {\bibinfo  {journal} {Phys. Rev.}\ }\textbf {\bibinfo {volume}
			{156}},\ \bibinfo {pages} {637} (\bibinfo {year} {1967})},\ \bibinfo {note}
	{the experiment found that $R_{yx}=43.1\times10^{-13}\ \Omega\cdot cm/G$ at
		room temperature, with $4\pi Ms=21.6\ kG$ and $\rho_{xx}=9.5\times 10^{-6}\
		\Omega\cdot cm$, the anomalous Hall angle will be
		$\theta^{AH}=\sigma_H\rho_{xx}=4\pi M_s R_{yx}/\rho_{xx}=0.0098$}\BibitemShut
	{NoStop}%
	\bibitem [{\citenamefont {Wang}\ \emph {et~al.}(2006)\citenamefont {Wang},
		\citenamefont {Yates}, \citenamefont {Souza},\ and\ \citenamefont
		{Vanderbilt}}]{PhysRevB.74.195118}%
	\BibitemOpen
	\bibfield  {author} {\bibinfo {author} {\bibfnamefont {X.}~\bibnamefont
			{Wang}}, \bibinfo {author} {\bibfnamefont {J.~R.}\ \bibnamefont {Yates}},
		\bibinfo {author} {\bibfnamefont {I.}~\bibnamefont {Souza}},\ and\ \bibinfo
		{author} {\bibfnamefont {D.}~\bibnamefont {Vanderbilt}},\ }\href
	{https://doi.org/10.1103/PhysRevB.74.195118} {\bibfield  {journal} {\bibinfo
			{journal} {Phys. Rev. B}\ }\textbf {\bibinfo {volume} {74}},\ \bibinfo
		{pages} {195118} (\bibinfo {year} {2006})}\BibitemShut {NoStop}%
	\bibitem [{\citenamefont {Kresse}\ and\ \citenamefont
		{Hafner}(1993)}]{PhysRevB.47.558}%
	\BibitemOpen
	\bibfield  {author} {\bibinfo {author} {\bibfnamefont {G.}~\bibnamefont
			{Kresse}}\ and\ \bibinfo {author} {\bibfnamefont {J.}~\bibnamefont
			{Hafner}},\ }\href {https://doi.org/10.1103/PhysRevB.47.558} {\bibfield
		{journal} {\bibinfo  {journal} {Phys. Rev. B}\ }\textbf {\bibinfo {volume}
			{47}},\ \bibinfo {pages} {558} (\bibinfo {year} {1993})}\BibitemShut
	{NoStop}%
	\bibitem [{\citenamefont {Kresse}\ and\ \citenamefont
		{Furthm\"uller}(1996)}]{PhysRevB.54.11169}%
	\BibitemOpen
	\bibfield  {author} {\bibinfo {author} {\bibfnamefont {G.}~\bibnamefont
			{Kresse}}\ and\ \bibinfo {author} {\bibfnamefont {J.}~\bibnamefont
			{Furthm\"uller}},\ }\href {https://doi.org/10.1103/PhysRevB.54.11169}
	{\bibfield  {journal} {\bibinfo  {journal} {Phys. Rev. B}\ }\textbf {\bibinfo
			{volume} {54}},\ \bibinfo {pages} {11169} (\bibinfo {year}
		{1996})}\BibitemShut {NoStop}%
	\bibitem [{\citenamefont {{Vosko}}\ \emph {et~al.}(1980)\citenamefont
		{{Vosko}}, \citenamefont {{Wilk}},\ and\ \citenamefont
		{{Nusair}}}]{1980CaJPh}%
	\BibitemOpen
	\bibfield  {author} {\bibinfo {author} {\bibfnamefont {S.~H.}\ \bibnamefont
			{{Vosko}}}, \bibinfo {author} {\bibfnamefont {L.}~\bibnamefont {{Wilk}}},\
		and\ \bibinfo {author} {\bibfnamefont {M.}~\bibnamefont {{Nusair}}},\ }\href
	{https://doi.org/10.1139/p80-159} {\bibfield  {journal} {\bibinfo  {journal}
			{Canadian Journal of Physics}\ }\textbf {\bibinfo {volume} {58}},\ \bibinfo
		{pages} {1200} (\bibinfo {year} {1980})}\BibitemShut {NoStop}%
	\bibitem [{\citenamefont {Bl\"ochl}(1994)}]{PhysRevB.50.17953}%
	\BibitemOpen
	\bibfield  {author} {\bibinfo {author} {\bibfnamefont {P.~E.}\ \bibnamefont
			{Bl\"ochl}},\ }\href {https://doi.org/10.1103/PhysRevB.50.17953} {\bibfield
		{journal} {\bibinfo  {journal} {Phys. Rev. B}\ }\textbf {\bibinfo {volume}
			{50}},\ \bibinfo {pages} {17953} (\bibinfo {year} {1994})}\BibitemShut
	{NoStop}%
	\bibitem [{\citenamefont {Kresse}\ and\ \citenamefont
		{Joubert}(1999)}]{PhysRevB.59.1758}%
	\BibitemOpen
	\bibfield  {author} {\bibinfo {author} {\bibfnamefont {G.}~\bibnamefont
			{Kresse}}\ and\ \bibinfo {author} {\bibfnamefont {D.}~\bibnamefont
			{Joubert}},\ }\href {https://doi.org/10.1103/PhysRevB.59.1758} {\bibfield
		{journal} {\bibinfo  {journal} {Phys. Rev. B}\ }\textbf {\bibinfo {volume}
			{59}},\ \bibinfo {pages} {1758} (\bibinfo {year} {1999})}\BibitemShut
	{NoStop}%
	\bibitem [{\citenamefont {Monkhorst}\ and\ \citenamefont
		{Pack}(1976)}]{PhysRevB.13.5188}%
	\BibitemOpen
	\bibfield  {author} {\bibinfo {author} {\bibfnamefont {H.~J.}\ \bibnamefont
			{Monkhorst}}\ and\ \bibinfo {author} {\bibfnamefont {J.~D.}\ \bibnamefont
			{Pack}},\ }\href {https://doi.org/10.1103/PhysRevB.13.5188} {\bibfield
		{journal} {\bibinfo  {journal} {Phys. Rev. B}\ }\textbf {\bibinfo {volume}
			{13}},\ \bibinfo {pages} {5188} (\bibinfo {year} {1976})}\BibitemShut
	{NoStop}%
	\bibitem [{\citenamefont {Thouless}\ \emph {et~al.}(1982)\citenamefont
		{Thouless}, \citenamefont {Kohmoto}, \citenamefont {Nightingale},\ and\
		\citenamefont {den Nijs}}]{PhysRevLett.49.405}%
	\BibitemOpen
	\bibfield  {author} {\bibinfo {author} {\bibfnamefont {D.~J.}\ \bibnamefont
			{Thouless}}, \bibinfo {author} {\bibfnamefont {M.}~\bibnamefont {Kohmoto}},
		\bibinfo {author} {\bibfnamefont {M.~P.}\ \bibnamefont {Nightingale}},\ and\
		\bibinfo {author} {\bibfnamefont {M.}~\bibnamefont {den Nijs}},\ }\href
	{https://doi.org/10.1103/PhysRevLett.49.405} {\bibfield  {journal} {\bibinfo
			{journal} {Phys. Rev. Lett.}\ }\textbf {\bibinfo {volume} {49}},\ \bibinfo
		{pages} {405} (\bibinfo {year} {1982})}\BibitemShut {NoStop}%
	\bibitem [{\citenamefont {Pizzi}\ \emph {et~al.}(2020)\citenamefont {Pizzi},
		\citenamefont {Vitale}, \citenamefont {Arita}, \citenamefont {Bl{\"{u}}gel},
		\citenamefont {Freimuth}, \citenamefont {G{\'{e}}ranton}, \citenamefont
		{Gibertini}, \citenamefont {Gresch}, \citenamefont {Johnson}, \citenamefont
		{Koretsune}, \citenamefont {Iba{\~{n}}ez-Azpiroz}, \citenamefont {Lee},
		\citenamefont {Lihm}, \citenamefont {Marchand}, \citenamefont {Marrazzo},
		\citenamefont {Mokrousov}, \citenamefont {Mustafa}, \citenamefont {Nohara},
		\citenamefont {Nomura}, \citenamefont {Paulatto}, \citenamefont
		{Ponc{\'{e}}}, \citenamefont {Ponweiser}, \citenamefont {Qiao}, \citenamefont
		{Th\"{o}le}, \citenamefont {Tsirkin}, \citenamefont {Wierzbowska},
		\citenamefont {Marzari}, \citenamefont {Vanderbilt}, \citenamefont {Souza},
		\citenamefont {Mostofi},\ and\ \citenamefont {Yates}}]{Pizzi_2020}%
	\BibitemOpen
	\bibfield  {author} {\bibinfo {author} {\bibfnamefont {G.}~\bibnamefont
			{Pizzi}}, \bibinfo {author} {\bibfnamefont {V.}~\bibnamefont {Vitale}},
		\bibinfo {author} {\bibfnamefont {R.}~\bibnamefont {Arita}}, \bibinfo
		{author} {\bibfnamefont {S.}~\bibnamefont {Bl{\"{u}}gel}}, \bibinfo {author}
		{\bibfnamefont {F.}~\bibnamefont {Freimuth}}, \bibinfo {author}
		{\bibfnamefont {G.}~\bibnamefont {G{\'{e}}ranton}}, \bibinfo {author}
		{\bibfnamefont {M.}~\bibnamefont {Gibertini}}, \bibinfo {author}
		{\bibfnamefont {D.}~\bibnamefont {Gresch}}, \bibinfo {author} {\bibfnamefont
			{C.}~\bibnamefont {Johnson}}, \bibinfo {author} {\bibfnamefont
			{T.}~\bibnamefont {Koretsune}}, \bibinfo {author} {\bibfnamefont
			{J.}~\bibnamefont {Iba{\~{n}}ez-Azpiroz}}, \bibinfo {author} {\bibfnamefont
			{H.}~\bibnamefont {Lee}}, \bibinfo {author} {\bibfnamefont {J.-M.}\
			\bibnamefont {Lihm}}, \bibinfo {author} {\bibfnamefont {D.}~\bibnamefont
			{Marchand}}, \bibinfo {author} {\bibfnamefont {A.}~\bibnamefont {Marrazzo}},
		\bibinfo {author} {\bibfnamefont {Y.}~\bibnamefont {Mokrousov}}, \bibinfo
		{author} {\bibfnamefont {J.~I.}\ \bibnamefont {Mustafa}}, \bibinfo {author}
		{\bibfnamefont {Y.}~\bibnamefont {Nohara}}, \bibinfo {author} {\bibfnamefont
			{Y.}~\bibnamefont {Nomura}}, \bibinfo {author} {\bibfnamefont
			{L.}~\bibnamefont {Paulatto}}, \bibinfo {author} {\bibfnamefont
			{S.}~\bibnamefont {Ponc{\'{e}}}}, \bibinfo {author} {\bibfnamefont
			{T.}~\bibnamefont {Ponweiser}}, \bibinfo {author} {\bibfnamefont
			{J.}~\bibnamefont {Qiao}}, \bibinfo {author} {\bibfnamefont {F.}~\bibnamefont
			{Th\"{o}le}}, \bibinfo {author} {\bibfnamefont {S.~S.}\ \bibnamefont
			{Tsirkin}}, \bibinfo {author} {\bibfnamefont {M.}~\bibnamefont
			{Wierzbowska}}, \bibinfo {author} {\bibfnamefont {N.}~\bibnamefont
			{Marzari}}, \bibinfo {author} {\bibfnamefont {D.}~\bibnamefont {Vanderbilt}},
		\bibinfo {author} {\bibfnamefont {I.}~\bibnamefont {Souza}}, \bibinfo
		{author} {\bibfnamefont {A.~A.}\ \bibnamefont {Mostofi}},\ and\ \bibinfo
		{author} {\bibfnamefont {J.~R.}\ \bibnamefont {Yates}},\ }\href
	{https://doi.org/10.1088/1361-648x/ab51ff} {\bibfield  {journal} {\bibinfo
			{journal} {Journal of Physics: Condensed Matter}\ }\textbf {\bibinfo {volume}
			{32}},\ \bibinfo {pages} {165902} (\bibinfo {year} {2020})}\BibitemShut
	{NoStop}%
	\bibitem [{\citenamefont {Tsirkin}(2020)}]{wannier-berri1}%
	\BibitemOpen
	\bibfield  {author} {\bibinfo {author} {\bibfnamefont {S.}~\bibnamefont
			{Tsirkin}},\ }\href {https://arxiv.org/abs/2008.07992} {\bibfield  {journal}
		{\bibinfo  {journal} {arXiv2008.07992}\ } (\bibinfo {year}
		{2020})}\BibitemShut {NoStop}%
	\bibitem [{\citenamefont {Destraz}\ \emph {et~al.}(2020)\citenamefont
		{Destraz}, \citenamefont {Das}, \citenamefont {Tsirkin}, \citenamefont {Xu},
		\citenamefont {Neupert}, \citenamefont {Chang}, \citenamefont {Schilling},
		\citenamefont {Grushin}, \citenamefont {Kohlbrecher}, \citenamefont {Keller},
		\citenamefont {Puphal}, \citenamefont {Pomjakushina},\ and\ \citenamefont
		{White}}]{wannier-berri2}%
	\BibitemOpen
	\bibfield  {author} {\bibinfo {author} {\bibfnamefont {D.}~\bibnamefont
			{Destraz}}, \bibinfo {author} {\bibfnamefont {L.}~\bibnamefont {Das}},
		\bibinfo {author} {\bibfnamefont {S.}~\bibnamefont {Tsirkin}}, \bibinfo
		{author} {\bibfnamefont {Y.}~\bibnamefont {Xu}}, \bibinfo {author}
		{\bibfnamefont {T.}~\bibnamefont {Neupert}}, \bibinfo {author} {\bibfnamefont
			{J.}~\bibnamefont {Chang}}, \bibinfo {author} {\bibfnamefont
			{A.}~\bibnamefont {Schilling}}, \bibinfo {author} {\bibfnamefont
			{A.}~\bibnamefont {Grushin}}, \bibinfo {author} {\bibfnamefont
			{J.}~\bibnamefont {Kohlbrecher}}, \bibinfo {author} {\bibfnamefont
			{L.}~\bibnamefont {Keller}}, \bibinfo {author} {\bibfnamefont
			{P.}~\bibnamefont {Puphal}}, \bibinfo {author} {\bibfnamefont
			{E.}~\bibnamefont {Pomjakushina}},\ and\ \bibinfo {author} {\bibfnamefont
			{J.}~\bibnamefont {White}},\ }\href
	{https://www.nature.com/articles/s41535-019-0207-7} {\bibfield  {journal}
		{\bibinfo  {journal} {npj Quantum Materials}\ }\textbf {\bibinfo {volume}
			{5}} (\bibinfo {year} {2020})}\BibitemShut {NoStop}%
	\bibitem [{\citenamefont {Marzari}\ and\ \citenamefont
		{Vanderbilt}(1997)}]{PhysRevB.56.12847}%
	\BibitemOpen
	\bibfield  {author} {\bibinfo {author} {\bibfnamefont {N.}~\bibnamefont
			{Marzari}}\ and\ \bibinfo {author} {\bibfnamefont {D.}~\bibnamefont
			{Vanderbilt}},\ }\href {https://doi.org/10.1103/PhysRevB.56.12847} {\bibfield
		{journal} {\bibinfo  {journal} {Phys. Rev. B}\ }\textbf {\bibinfo {volume}
			{56}},\ \bibinfo {pages} {12847} (\bibinfo {year} {1997})}\BibitemShut
	{NoStop}%
	\bibitem [{\citenamefont {Souza}\ \emph {et~al.}(2001)\citenamefont {Souza},
		\citenamefont {Marzari},\ and\ \citenamefont
		{Vanderbilt}}]{PhysRevB.65.035109}%
	\BibitemOpen
	\bibfield  {author} {\bibinfo {author} {\bibfnamefont {I.}~\bibnamefont
			{Souza}}, \bibinfo {author} {\bibfnamefont {N.}~\bibnamefont {Marzari}},\
		and\ \bibinfo {author} {\bibfnamefont {D.}~\bibnamefont {Vanderbilt}},\
	}\href {https://doi.org/10.1103/PhysRevB.65.035109} {\bibfield  {journal}
		{\bibinfo  {journal} {Phys. Rev. B}\ }\textbf {\bibinfo {volume} {65}},\
		\bibinfo {pages} {035109} (\bibinfo {year} {2001})}\BibitemShut {NoStop}%
	\bibitem [{\citenamefont {Marzari}\ \emph {et~al.}(2012)\citenamefont
		{Marzari}, \citenamefont {Mostofi}, \citenamefont {Yates}, \citenamefont
		{Souza},\ and\ \citenamefont {Vanderbilt}}]{RevModPhys.84.1419}%
	\BibitemOpen
	\bibfield  {author} {\bibinfo {author} {\bibfnamefont {N.}~\bibnamefont
			{Marzari}}, \bibinfo {author} {\bibfnamefont {A.~A.}\ \bibnamefont
			{Mostofi}}, \bibinfo {author} {\bibfnamefont {J.~R.}\ \bibnamefont {Yates}},
		\bibinfo {author} {\bibfnamefont {I.}~\bibnamefont {Souza}},\ and\ \bibinfo
		{author} {\bibfnamefont {D.}~\bibnamefont {Vanderbilt}},\ }\href
	{https://doi.org/10.1103/RevModPhys.84.1419} {\bibfield  {journal} {\bibinfo
			{journal} {Rev. Mod. Phys.}\ }\textbf {\bibinfo {volume} {84}},\ \bibinfo
		{pages} {1419} (\bibinfo {year} {2012})}\BibitemShut {NoStop}%
\end{thebibliography}

%

\end{document}